\documentclass{cccg20}
\usepackage[utf8]{inputenc}
\usepackage{graphicx,amssymb,amsmath}
\usepackage[appendix=inline]{apxproof}

\usepackage{subcaption}
\usepackage{multicol}
\usepackage{enumitem}
\usepackage{hyperref}
\hypersetup{hidelinks}
 
\usepackage{algorithm}
\usepackage{algpseudocode}

\algdef{SE}[DOWHILE]{Do}{DoWhile}{\algorithmicdo}[1]{\algorithmicwhile\ #1}

\newtheoremstyle{mytheoremstyle} 
    {\topsep} 
    {\topsep} 
    {\em} 
    {} 
    {\bf} 
    {} 
    { } 
    {} 

\theoremstyle{mytheoremstyle}

\newcommand{\Subsection}[1]{~~\\\noindent{\bf #1.}~~~}

\newtheoremrep{definition}{Definition}
\newtheoremrep{problem}{Problem}

\newtheoremrep{theorem}{Theorem}
\newtheoremrep{invar}{Invariant}
\newtheoremrep{lemma}[theorem]{Lemma}
\newtheoremrep{cor}[theorem]{Corollary}
\newtheoremrep{obs}{Observation}
\newtheoremrep{conj}{Conjecture}
\newtheoremrep{prop}[theorem]{Proposition}

\newcommand{\esim}{\ensuremath{\overset{*}{\sim}}}

\makeatletter
\let\OldStatex\Statex
\renewcommand{\Statex}[1][3]{%
  \setlength\@tempdima{\algorithmicindent}%
  \OldStatex\hskip\dimexpr#1\@tempdima\relax}
\makeatother

\title{Minimizing The Maximum Distance Traveled To Form Patterns With Systems of Mobile Robots}

\author{Jared Coleman\thanks{California State University,
        Long Beach, {\tt jared.coleman@student.csulb.edu}}
        \and
        Evangelos Kranakis\thanks{Carleton University,
        School of Computer Science, Ottawa, Ontario K1S 5B6, Canada,
        Research supported in part by NSERC Discovery grant
        {\tt kranakis@scs.carleton.ca}}
        \and
        Oscar Morales-Ponce\thanks{California State University,
        Long Beach, {\tt  oscar.morales-ponce@csulb.edu}}
        \and
        Jaroslav Opatrny\thanks{Concordia University, 
        Department Computer Science and Engineering, Montréal QC H3G 1M8 Canada
        {\tt opatrny@cs.concordia.ca}}
        \and
        Jorge Urrutia\thanks{Instituto de Matematicas, UNAM, 
        Mexico City, Mexico
        {\tt urrutia@matem.unam.mx}}
        \and
        Birgit Vogtenhuber\thanks{Institute of Software Technology, 
        University of Technology, Graz, Austria
        {\tt bvogt@ist.tugraz.at}}}

\index{Coleman, Jared}
\index{Kranakis, Evangelos}
\index{Morales Ponce, Oscar}
\index{Opatrny, Jaroslav}
\index{Urrutia, Jorge}
\index{Vogtenhuber, Birgit}

\begin{document}
\thispagestyle{empty}
\maketitle

\begin{abstract}
In the pattern formation problem, robots in a system must self-coordinate to form a given pattern, regardless of translation, rotation, uniform-scaling, and/or reflection.
In other words, a valid final configuration of the system is a formation that is \textit{similar} to the desired pattern.
While there has been no shortage of research in the pattern formation problem under a variety of assumptions, models, and contexts, we consider the additional constraint that the maximum distance traveled among all robots in the system is minimum.
Existing work in pattern formation and closely related problems are typically application-specific or not concerned with optimality (but rather feasibility).
We show the necessary conditions any optimal solution must satisfy and present a solution for systems of three robots.
Our work also led to an interesting result that has applications beyond pattern formation.
Namely, a metric for comparing two triangles where a distance of $0$ indicates the triangles are similar, and $1$ indicates they are \emph{fully dissimilar}.
\end{abstract}

\section{Introduction}
\label{sec:intro}
While distributed systems have clear advantages over centralized ones, their complexity has stunted their potential in the mobile robotics market.
Where distributed systems are cheap to build, scalable, and fault-tolerant in theory, they are extremely difficult to properly design in practice~\cite{schetter2003multiple}.
In this paper, we present results from a study on pattern formation, a common problem in distributed robotics. 
In the pattern formation problem, a system of mobile robots on the plane must move to form a given pattern.
While this problem has been studied extensively, we consider the additional constraint that the maximum distance traveled among all robots must be minimum.
For the purpose of this paper, we call solutions that satisfy this constraint \textit{optimal}.

The main goal of this study is to develop a theoretical understanding of the pattern formation problem.
In this study, we make contributions to establishing this baseline and, in doing so, make many interesting observations about properties and limitations for patterns and  the systems that form them.

\Subsection{Our Contributions}
The goal of this study is to develop a theoretical understanding of the min-max traversal pattern formation problem.
To do so, we first explore the necessary conditions that any optimal solution must satisfy.
For example, we prove in Section~\ref{sec:necessary_conditions} (Lemma \ref{lemma:minimum_3_robots}) that for any optimal solution, at least three robots must travel exactly the maximum distance.
Notice that for systems of three robots, this means all three robots must move exactly the same distance, regardless of the pattern they must form.
Clearly, the three-robot case is an important lower bound for the general case and is therefore the primary focus of this study.
In Section~\ref{sec:main_results}, we present an algorithm for computing the optimal solution for systems of three robots.
While not directly applicable, the three-robot solution has important implications on systems of many robots.
In Section~\ref{sec:conclusion}, we discuss these implications in further detail.

Our work on systems of three robots also yielded a surprising, but profound result.
In Section~\ref{sec:metric}, we prove that by modifying the aforementioned algorithm slightly, we can use it as a metric for measuring the similarity between two arbitrary triangles.
This has potential applications beyond pattern formation for mobile robotic systems, like computational geometry and computer vision.

\Subsection{Models}\label{sec:models}
In this paper, we are interested in the globally optimal solution to the pattern formation problem.
Different models, however, may or may not be able to compute the optimal solution.
In this section, we briefly discuss various models used in related literature and their implications on the pattern formation problem.
All models discussed in this paper follow the \textit{look, compute, move} execution cycle.
In the \textit{look} phase, each robot observes the position of all other robots in the system (either globally or relative to their own local coordinate frame).
Then, robots \textit{compute} a solution and \textit{move} some distance towards it.
We also assume that, in each cycle, all robots move the same distance $\delta$ toward their destination unless they reach it, in which case they move some distance less than $\delta$.

In accordance with related literature, we consider whether robots in the system are globally coordinated, oblivious, oriented, and/or synchronous. 
Robots are \textit{globally coordinated} if they have access to a global coordinate system, otherwise they are said to be \textit{locally coordinated}.
Robots are \textit{oblivious} if they do not have access to previous states of the system.
In oblivious models, a solution must be computed using only a snapshot of the system at a given time.
Robots are \textit{oriented} if they have a common sense of direction (i.e. North, South, East, and West), otherwise they are \textit{unoriented}.
Robots are synchronous if they start each phase of their \textit{look, compute, move} cycles at the same time (according to some global clock).
In this paper, we also assume synchronous robots move at the same speed.

It has been shown that asynchronous and oblivious robots cannot form any arbitrary pattern (Theorem 3.1 in \cite{suzuki1999distributed}).
It has also been shown that locally coordinated, synchronous robots cannot form any arbitrary pattern (even sub-optimally) \cite{suzuki1999distributed}, but that locally coordinated, asynchronous robots can as long as they are oriented \cite{flocchini2008arbitrary} due to possible symmetry in the initial configuration of robots.
We assume robots are in general position and therefore do not consider the special case where robots are symmetric.
Note that for any special case where robots are synchronous with each other, we can perturbate each robot's position by some small arbitrarily random amount to break symmetry.
Table~\ref{table:models} is a summary of which models can and cannot form patterns optimally or sub-optimally for systems of three robots.

\begin{table*}[h!]
   \centering
   \small
   \begin{tabular}{| c | c | c | c || c r |} 
   \hline
   \textbf{Globally} & \textbf{Oblivious} & \textbf{Synchronous} & \textbf{Oriented} & \multicolumn{2}{|c|}{\textbf{Pattern}} \\ 
   \textbf{Coordinated} & & & & \multicolumn{2}{|c|}{\textbf{Formable}} \\ \hline
   Yes & -   & Yes & -   & Optimal    & Corollary~\ref{cor:solution},Theorem~\ref{theorem:oblivious}    \\ \cline{2-6}
       & No  & No  & -   & Valid      & \cite{suzuki1999distributed}      \\ \hline
   No  & Yes & Yes & Yes & Optimal    & Corollary~\ref{cor:solution},Theorem~\ref{theorem:oblivious}    \\ \cline{3-6}
       &     & -   & No  & Impossible & \cite{suzuki1999distributed}      \\ \cline{2-6}
       & No  & Yes & -   & Optimal    & Corollary~\ref{cor:solution}    \\ \cline{3-6}
       &     & No  & Yes & Valid      & \cite{flocchini2008arbitrary}     \\ \cline{4-6}
       &     &     & No  & Impossible & \cite{flocchini2008arbitrary}     \\ \hline
   \end{tabular}
   \caption[Comparison of models]{
      A globally optimal pattern is only formable in the general case under some models.
      Under some models, a valid sub-optimal formation can always be formed while under others, valid formations are not formable \textit{at all} in the general case.
      Note that the results reported in this paper are only valid for systems of three robots.
   }
   \label{table:models}
\end{table*}

In this paper, we show that our solution for systems of three robots is valid under all globally coordinated, synchronous models and under the locally coordinated, oblivious, synchronous, and oriented model.

\Subsection{Notation}
For any system of $n$ robots, we denote their initial positions by
$R = (r_0, r_1, \ldots, r_{n-1})$ (robot $i$ is at position $r_i$). 
We define a pattern to be a sequence of distinct points on the plane and use capital letters, like $P$ and $S$, to denote them. 
Lower-case letters and subscript indices are used to denote the elements of the sequence. 
For example, $p_i$ is the $i^{th}$ element of $P$. 
Sets of sequences of distinct points on the plane (e.g. sets of patterns) are denoted in calligraphic font, for example $\mathcal{P}$ and $\mathcal{S}$. 
Elements of these sets are denoted with their non-calligraphic equivalent and a superscript index. 
For example, $S^i$ is the $i^{th}$ element of $\mathcal{S}$ and $s^i_j$ is the $j^{th}$ element of $S^i$. 

The number of elements in a sequence $P$, or its length, is denoted by $|P|$.
Two sequences $P$ and $Q$ are equivalent, or $P=Q$, if and only if $|P|=|Q|$ and $p_i=q_i$ for $0 \leq i < |P|$.
We say $P$ and $Q$ are similar, or $P \sim Q$ if and only if there exists some translation, rotation, uniform scaling, and/or reflection of any permutation of $P$ that is equivalent to $Q$.
$P$ and $Q$ are rigidly similar, or $P \esim Q$ if and only if there exists some translation, rotation, and/or uniform scaling of $P$, say $P^\prime$, such that $P^\prime = Q$.
Observe that $P \esim Q \Rightarrow P \sim Q$, but $P \sim Q \not\Rightarrow P \esim Q$.

Let $C(p, r)$ be the circle centered at $p$ with radius~$r$ and $D(p,r)$ be the closed disk with center~$p$ and radius~$r$.
Also, let $d(u, v)$ be the Euclidian distance between points~$u$ and~$v$. 

\Subsection{Outline}
This paper is organized as follows.
First, we formally introduce the problem statement in Section~\ref{sec:problem_statement} and discuss related work in Section~\ref{sec:related_work}.
Then, we discuss the necessary conditions any optimal solution must satisfy in Section~\ref{sec:necessary_conditions}.
In Section~\ref{sec:replication}, we introduce Replication, a tool we use in Section~\ref{sec:main_results} to show that our main contribution, an optimal solution for systems of three robots, is in fact optimal.
In Section~\ref{sec:metric}, we present a metric based on the optimal solution for systems of three robots.
In Section~\ref{sec:properties}, we discuss some properties of systems of three robots and the patterns they can form.
Finally, Section~\ref{sec:conclusion} concludes this study with a discussion about future work and the significance of our contributions toward a theoretical understanding of the pattern formation problem.

\section{Problem Statement}
\label{sec:problem_statement}
Consider a system of $n$ robots with initial positions
$R = (r_0, r_1, ..., r_{n-1})$.
The trajectory of robot~$i$ is defined as a continuous function $f_i(t)$  for all $t \geq 0$. 
A strategy $A$ defines a trajectory for every robot. 
Given a pattern $P$, we say that the strategy $A$ is \emph{valid} if there exists a time $t$
such that the robots' positions are similar to $P$.
Otherwise the strategy is \emph{invalid}. 
To simplify notation we say robots that use a valid strategy \textit{form}~$P$.
Let $t(A)$ be the earliest time at which the robots form $P$ using strategy $A$.
The distance that each robot traverses is defined as 
$d^A_i = \int_0^{t(A)}f^A_{i}(t) dt$.

In this study we are interested in a strategy that minimizes the maximum distance any robot traverses to form the desired pattern: 

\begin{problem}[Min-Max Traversal Pattern Formation]
\label{prob:MinMaxPF}
Given a system of $n \geq 3$ robots with initial positions $R$ and a pattern $P$, determine the minimum $d^*$ for which there exists a valid strategy for forming $P$ such that every robot travels at most distance $d^*$.
Formally:
$$
   d^* = \min_{\forall A \in \mathcal{A}}(\max_{0 \leq i < n}(d^A_i))
$$
where $\mathcal{A}$ is the set of all valid strategies.
\end{problem}


\section{Related Work}
\label{sec:related_work}
The pattern formation problem has been studied extensively under a variety of assumptions, models, and contexts.
Many researchers use the pattern formation problem to study the algorithmic limitations of autonomous mobile robots \cite{flocchini2008arbitrary, suzuki1999distributed}.
It has been shown, for example, that systems of synchronous robots with initially symmetric positions cannot form any geometric pattern \cite{suzuki1999distributed} but that systems of asynchronous robots with compasses (A global sense of North/South and East/West) can \cite{flocchini2008arbitrary}.
We mitigate the problems that symmetry introduces to the pattern formation problem by assuming robots initial positions are random, and the probability of exact symmetry approaches zero.
Since we are interested in finding any \textit{theoretically} optimal solution for the general case, a feasibility discussion is out of the scope for this paper and left as future work.

Researchers have proposed solutions for many variants of the pattern formation problem.
For example, it has been shown that it is possible to form a uniform circle (one such that the distance between neighboring robots on the circle is equal) for any system of robots arbitrarily deployed on the plane \cite{flocchini2014distributed}.
Other variations of the pattern formation that have been studied include gathering on a ring \cite{klasing2008gathering},
point-convergence \cite{cohen2005convergence},
and forming a series of patterns in succession \cite{das2010computational}.
There has also been work in variations of these problems where robots have visibility constraints, that is, they can only see other robots in the system if they are within a given distance \cite{ando1999distributed,flocchini2005gathering,cohen2008local}.
Various methods and solutions for bio-inspired pattern formation are reviewed in \cite{oh2017bio}.
When the destination positions are known, the pattern formation is reduced to robot-destination matching.
There are many available solutions for these kinds of variants of the problem that guarantee a variety of different properties (i.e. no collision, minimum total distance traveled, etc.) \cite{alonso-mora2011multi}.
Solutions typically involve a combinatorial optimization algorithm for the assignment problem, like the Hungarian Algorithm~\cite{kuhn1955hungarian}.
The quantity and variety of the literature reflects the seemingly unlimited variants and applications of the pattern formation problem.
There is, however, no unifying theory that ties all these solutions together.
In this study, we make progress toward addressing this shortcoming of the field.

The pattern formation problem has also been studied from an operations research perspective. 
A solution has been proposed that formulates the problem as a second-order cone program \cite{derenick2007convex} and uses interior-point methods to solve it. 
This solution, however, relies on a prescribed assignment and does not consider reflection.
The authors report a constant runtime, but this is in the number of iterations of the convex optimization step, and does not consider the time to create the necessary data structures.
Our implementation has a time-complexity of $O(n^3)$ where $n$ is the number of robots in the system.
Some work has been done to incorporate assignment as well, but current solutions exist only for minimizing the \textit{total} distance traveled by all robots (as opposed to the maximum distance traveled by any robot in the system) \cite{agarwal2018simultaneous}.
While these solutions are practical and useful for many situations, they are not analytical and do not provide any insight into the properties of optimal solutions.
In this study, we develop a theoretical understanding of the pattern formation problem and work toward an analytical solution to the problem.

\section{Necessary Conditions}
\label{sec:necessary_conditions}
First, we start by characterizing an optimal solution.
In this section we present the necessary conditions that every optimal solution must satisfy. 
 
\Subsection{Critical Robots}
Throughout the paper, we use \textit{critical robots} to refer to robots which move the maximum distance (the solution).
In Lemma~\ref{lemma:minimum_3_robots} we show that in any optimal solution there are at least three critical robots.
\begin{lemmarep}
   \label{lemma:minimum_3_robots}
   Given a system of $n$ robots with initial positions $R=(r_0, r_1, \ldots, r_{n-1})$,
   let $d^*$ be the optimal solution for forming some pattern.
   Then at least three robots traverse exactly distance $d^*$. 
\end{lemmarep} 

\begin{appendixproof}
   Consider a valid strategy that yields the final robot destinations
   $F = (f_0, f_1, ..., f_{n-1})$ (i.e. robot $i$ moves from $r_i$ to $f_i$).
   Suppose for sake of contradiction that less than three robots traverse distance $d^*$ to reach their final destination.
   Therefore, $F$ has at most two vertices on the circumference of distinct disks with centers at $R$ and radius $d^*$ (Figure~\ref{fig:minimum_3_robots}).
   It is obvious that at least one robot must traverse distance $d^*$, otherwise $d^*$ is not optimal. We consider two cases:

   \textbf{Case 1:} Exactly $1$ robot traverses $d^*$ (Figure~\ref{fig:minimum_3_robots_1}). 
   Suppose, without loss of generality, that robot $0$  traverses exactly $d^*$. 
   For every  $i \neq 0$, let $D(f_i, l_i)$ be the largest circle totally contained in $D(r_i, d^*)$.
   Consider the pattern $F^\prime$ obtained by translating $F$ some non-zero distance
   less than $\underset{i \not= 0}{\min}(l_i)$ in the direction such that $f^\prime_0$ is some positive distance $\Delta d \leq \underset{i \not= 0}{\min}(l_i)$ closer to $r_0$ than $f_0$.
   The maximum distance any robot must travel to reach $f^\prime_i$ is 
   $d^* - \Delta d$, which contradicts the optimality of $d^*$. 

   \begin{figure*}[ht!]
      \centering
      \begin{subfigure}[t]{0.49\textwidth}
         \centering
         \includegraphics[width=\textwidth]{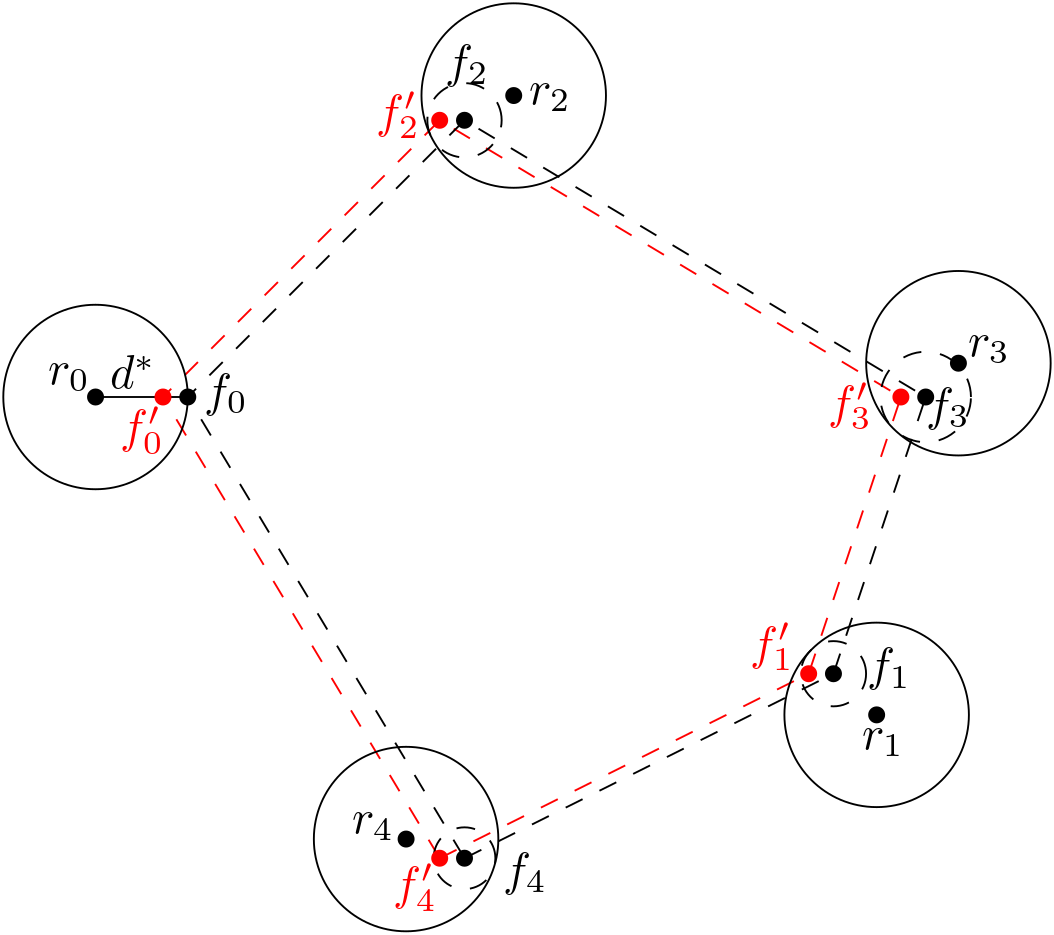}
         \caption{One robot traverses distance $d^*$.}
         \label{fig:minimum_3_robots_1}
      \end{subfigure}
      ~ 
      \begin{subfigure}[t]{0.49\textwidth}
         \centering
         \includegraphics[width=\textwidth]{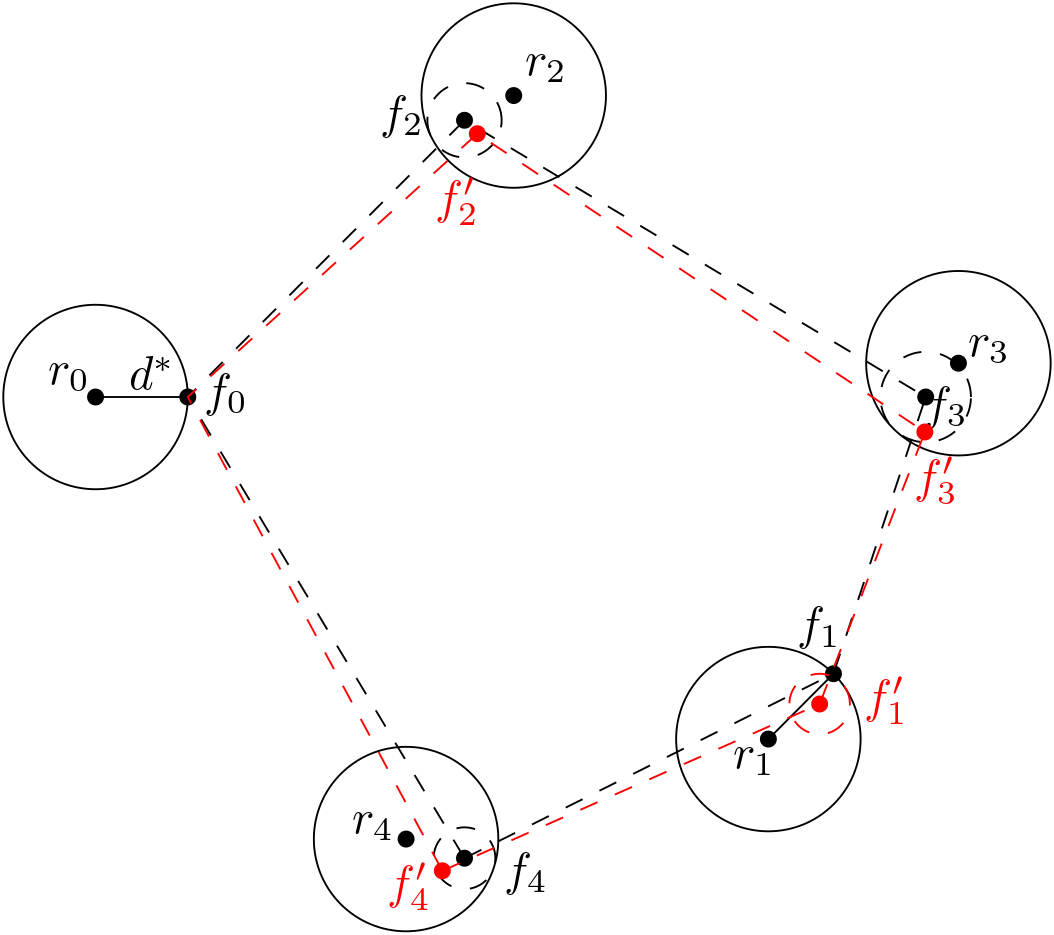}
         \caption{Two robots traverse distance $d^*$.}
         \label{fig:minimum_3_robots_2}
      \end{subfigure}
      \caption[Minimum of three critical robots]{
         When less than three robots traverse exactly $d^*$ and all other robots traverse some distance less than $d^*$, there is always a better solution obtainable by rotating, translating, and/or uniformly scaling the final destination points.
      }
      \label{fig:minimum_3_robots}
   \end{figure*}

   \textbf{Case 2:} Exactly $2$ robots traverse $d^*$ (Figure~\ref{fig:minimum_3_robots_2}). 
   Suppose, without loss of generality, that robots $0$ and $1$ traverse exactly $d^*$. 
   For every $i \neq 0, 1$, let $D(f_i, l_i)$ be the largest disk totally contained in $D(r_i, d^*)$.
   Consider a pattern $F^\prime \sim F$ obtained by fixing $f_0$ and rotating and/or uniformly scaling $F$ such that $f^\prime_1$ is closer to $r_1$ than $f_1$.
   Note that this is always possible and rotation is only necessary
   when $\overleftrightarrow{f_0 f_1}$ is tangent to $D(f_1, l_1)$.
   In this case, it is clear that $f^\prime$ can be obtained by rotating $F$ toward $D(f_1, l_1)$.
   Furthermore, since robots $i \neq 0, 1$ can be displaced some $\underset{i \not= 0}{\min}(l_i) > 0$ in any direction, it is always possible to obtain a $F^\prime$ such that only robot $0$ moves exactly $d^*$ and all others must traverse less than $d^*$.
   This reduces to Case 1 and thus is contradictory to the optimality of $d^*$.
\end{appendixproof}

Lemma~\ref{lemma:minimum_3_robots} does not prove the existence of any upper bound on the number of robots that move distance $d^*$.

\section{Replication}
\label{sec:replication}
In this section we present \emph{Replication} as a tool that we use to derive results presented later in the study.
The replication machine is based on pure geometry and resembles a Pantograph.
While replication is naturally applicable for any pattern with three or more vertices, we present replication for triangles in this study to simplify notation and proofs. 

\begin{definition}[Trivial Replication]
   The Trivial Replication of a triangle $P$ on a pair of points $(u, v)$ is the triangle rigidly similar to $P$ whose first two points are fixed to $u$ and $v$.
   Formally:
   \begin{center}
      $R_{Triv}(P, u, v) = T \esim P$ such that $t_0=u$, and $t_1=v$.
   \end{center}
\end{definition}

For any Trivial Replication $T=R_{Triv}(P, u, v)$, 
we call $u=t_0$ and $v=t_1$ its anchors. 
We call the third point, $t_2$, the \textit{Trivial Replication Point}.
Note that the Trivial Replication Point is not explicitly fixed to a prescribed point, rather, its position is entirely dependent on the triangle being replicated and the two anchors.

\begin{definition}[Replication Machine]
   The Replication Machine of a triangle $P$ on a point and a circle 
   $(u, C(v, r))$ is the infinite set of triangles rigidly similar to 
   $P$ whose first point is fixed to $u$ and whose second point is on the circle 
   $C(v, r)$.
   Formally:

   \begin{center}
      $R_{Mach}(P, u, v, r) = \{T \esim P \vert t_0=u, t_1 \in C(v, r) \}$.
   \end{center}

   or equivalently:

   \begin{center}
      $R_{Mach}(P, u, v, r) = \{R_{Triv}(P, u, v^\prime) \vert v^\prime \in C(v, r) \}$.
   \end{center}
\end{definition}

Observe that $R_{Mach}(P, u, v, r)$ is the set of all patterns rigidly similar to 
$P$ such that, for any $T \in R_{Mach}(P, u, v, r)$, $t_0$ is fixed to $u$ and $t_1$ is exactly distance $r$ from $v$.
Observe that each triangle in a Replication Machine is also a Trivial Replication of the same triangle.
We call the set of Trivial Replication Points of the Trivial Replications in
a Replication Machine \textit{Replication Machine Points}.

\begin{definition}[Replication Spanner]
   The Replication Spanner of a triangle $P$ on a pair of circles 
   $(C(u, r), C(v, r))$ is the infinite set of triangles rigidly similar to 
   $P$ whose first and second points are on the circles 
   $C(u, r)$ and $C(v, r)$, respectively.
   Formally:

   \begin{center}
      $R_{Span}(P, u, v, r) = \{T \esim P \vert t_0 \in C(u, r), t_1 \in C(v, r) \}$.
   \end{center}

   or equivalently:

   \begin{center}
      $R_{Span}(P, u, v, r) = \underset{u^\prime \in C(u, r)}{\bigcup} R_{Mach}(P, u^\prime, v, r)$.
   \end{center}
\end{definition}

$R_{Span}(P, u, v, r)$ is the set of all patterns rigidly similar to $P$ such that, for any 
$T \in R_{Span}(P, u, v, r)$, both $t_0$ and $t_1$ are exactly distance $r$ from $p$ and $q$, respectively.
We call the set of Trivial Replication Points of the Trivial Replications in a Replication Spanner \textit{Replication Spanner Points}.

It is starting to become clear why Replication is a useful tool for pattern formation.
Suppose $u$ and $v$ are the initial positions of two robots in a system that must form a triangle $P$.
Then $R_{Span}(P, u, v, r)$ is the set of all patterns rigidly similar to $T$ that the robots can form by each moving distance $r$.
Since we are dealing with a system of three robots (forming triangular patterns),
we know that all three robots are \textit{critical} (Lemma~\ref{lemma:minimum_3_robots}).
Therefore, the optimal pattern (without considering permutation or reflection) must be one from $R_{Span}(P, u, v, r)$ for some value of $r$.

\begin{lemmarep}
   Let $c$ be the Trivial Replication Point of a triangle $P$ on a pair of points 
   $(u, v)$. Then the set Replication Machine Points of $P$ on $(u, C(v, r))$ 
   is enclosed by the circle $C\left(c, r\frac{d(u, c)}{d(u, v)} \right)$.
\end{lemmarep}

\begin{proofsketch}
   First, we show that the Replication Machine Points form a circle.
   Consider the Trivial Replication $T = R_{Triv}(P, u, v)$ (note that $t_2 = c$) 
   and an arbitrary Trivial Replication $M \in R_{Mach}(P, u, v, r)$ 
   (note that $m_2$ is in the Trivial Replication Circle of $R_{Mach}(P, u, v, r)$).
   First, observe that since $T$ is rigidly similar to $M$, then 
   $d(u, m_1) = k ~ d(u, v)$ and 
   $d(u, m_2) = k ~ d(u, c)$ for some $k$, thus
   $\triangle u m_2 c$ is similar to 
   $\triangle u m_1 v$ and $d(c, m_2)$ must be proportional to $r$.

   In order to simplify the calculation of the circle's radius,
   consider the Trivial Replication $M \in R_{Mach}(P, u, v, r)$ such that $m_1$ is colinear with the line $\overleftrightarrow{uv}$.
   Observe that $k ~ d(u, v) = d(u, v) + r$ and 
   $k ~ d(u, c) = d(u, c) + d(c, m_2)$.
   Solving the system of equations results in $d(c, m_2) = r \frac{d(u, c)}{d(u, v)}$.
\end{proofsketch}
   
\begin{appendixproof}
   \begin{figure*}[ht!]
      \centering
      \begin{subfigure}[t]{0.49\textwidth}
         \centering
         \includegraphics[width=\textwidth]{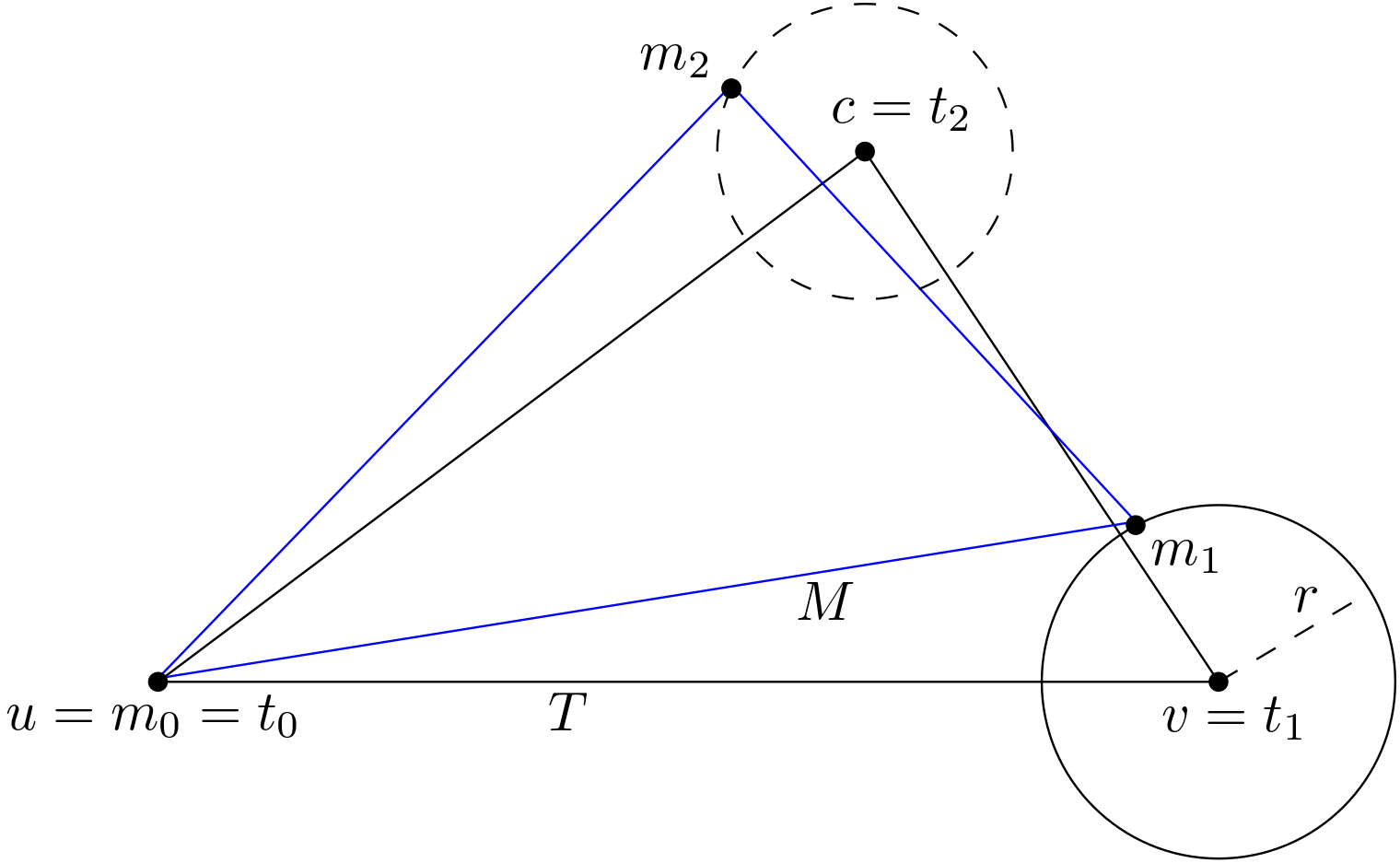}
         \caption{
            $M$ is one trivial replication in the infinite set $R_{Mach}(P, u, v, r)$.
         }
         \label{fig:replication_machine_one}
      \end{subfigure}
      ~ 
      \begin{subfigure}[t]{0.49\textwidth}
         \centering
         \includegraphics[width=\textwidth]{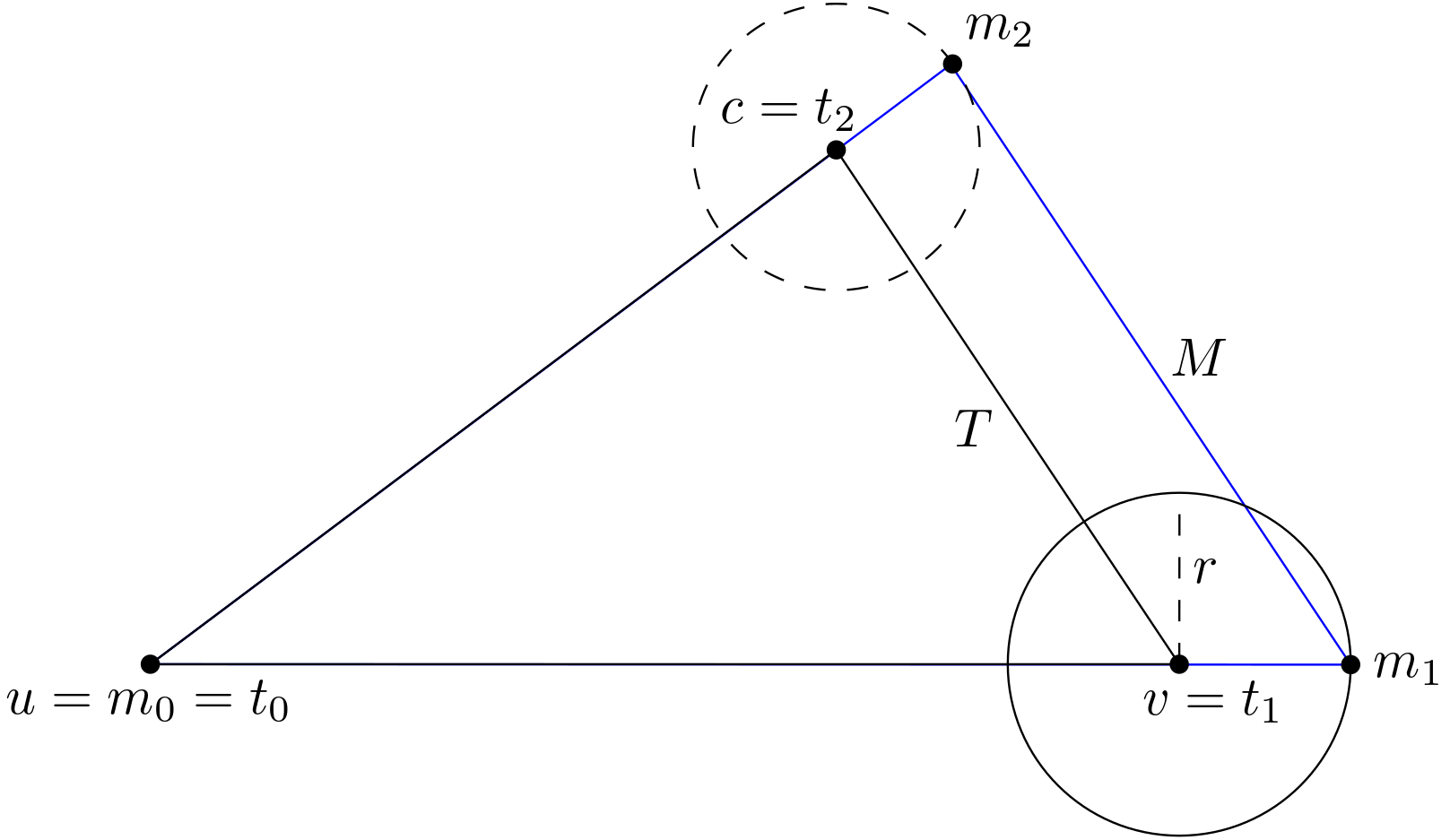}
         \caption{To simplify the calculation of the radius of the Replication Machine Circle, consider $M \in R_{Mach}(P, u, v, r)$ such that 
         $\overleftrightarrow{m_0 m_1}$ is colinear with $\overleftrightarrow{uv}$}.
         \label{fig:replication_machine_radius}
      \end{subfigure}
      \caption[Replication Machine Circle]{
         Replications for the Replication Machine $R_{Mach}(P, u, v, r)$.
      }
      \label{fig:replication_machine}
   \end{figure*}

   First, we show that Replication Machine Points form a circle.
   Consider the Trivial Replication $T = R_{Triv}(P, u, v)$ (note that $t_2 = c$) 
   and an arbitrary Trivial Replication $M \in R_{Mach}(P, u, v, r)$ 
   (note that $m_2$ is in the Trivial Replication Circle of $R_{Mach}(P, u, v, r)$).
   First, observe that since $T$ is rigidly similar to $M$, then 
   $d(u, m_1) = k ~ d(u, v)$ and 
   $d(u, m_2) = k ~ d(u, c)$ for some $k$, thus
   $\triangle u m_2 c$ is similar to 
   $\triangle u m_1 v$ and $d(c, m_2)$ must be proportional to $r$.
   (Figure~\ref{fig:replication_machine_one}).

   In order to simplify the calculation of the circle's radius,
   consider the Trivial Replication $M \in R_{Mach}(P, u, v, r)$ such that $m_1$
   is colinear with the line $\overleftrightarrow{uv}$ 
   (Figure~\ref{fig:replication_machine_radius}).
   Observe that $k ~ d(u, v) = d(u, v) + r$ and 
   $k ~ d(u, c) = d(u, c) + d(c, m_2)$.
   Solving the system of equations:
   \begin{align*}
      \frac{d(u, c) + d(c, m_2)}{d(u, c)} &= \frac{d(u, v) + r}{d(u, v)} \\
      d(c, m_2) &= d(u, c) ~ \frac{d(u, v) + r}{d(u, v)} - d(u, c) \\
      d(c, m_2) &= d(u, c) ~ \left( \frac{d(u, v) + r}{d(u, v)} - 1 \right) \\
      d(c, m_2) &= r \frac{d(u, c)}{d(u, v)}
   \end{align*}
\end{appendixproof}

We call $C \left(c, r \frac{d(u, c)}{d(u, v)} \right)$ the Replication Machine Circle of $R_{Mach}(P, u, v, r)$.

\begin{lemmarep}
   If $c$ is the Trivial Replication Point of a triangle $P$ 
   on a pair of points $(u, v)$. 
   Then $C\left(c, r\frac{d(u, c) + d(v, c)}{d(u, v)} \right)$
   is the smallest circle that encloses the Replication Spanner Points of
   $P$ on $(C(u, r), C(v, r))$.
\end{lemmarep}

\begin{proofsketch}
   Consider the Trivial Replication $T = R_{Triv}(P, u, v)$, and
   the replication machines $\mathcal{M} = R_{Mach}((p_1, p_0, p_2), v, u, r)$
   and, for some $M \in \mathcal{M}$, $\mathcal{S} = R_{Mach}(P, m_0, v, r)$.
   
   Observe that for any $S \in \mathcal{S}$, by the definition of Replication Spanner,
   $S \in R_{Span}(P, u, v, r)$.
   Observe that the center of the Replication Machine Circle of $\mathcal{S}$ is in 
   the Replication Machine Circle of $\mathcal{M}$.
   Therefore, $c$ is the center-of-centers of two Replication Machine Circles.
\end{proofsketch}

\begin{appendixproof}
   \begin{figure}[!ht]
      \centering 
      \includegraphics[width=\columnwidth]{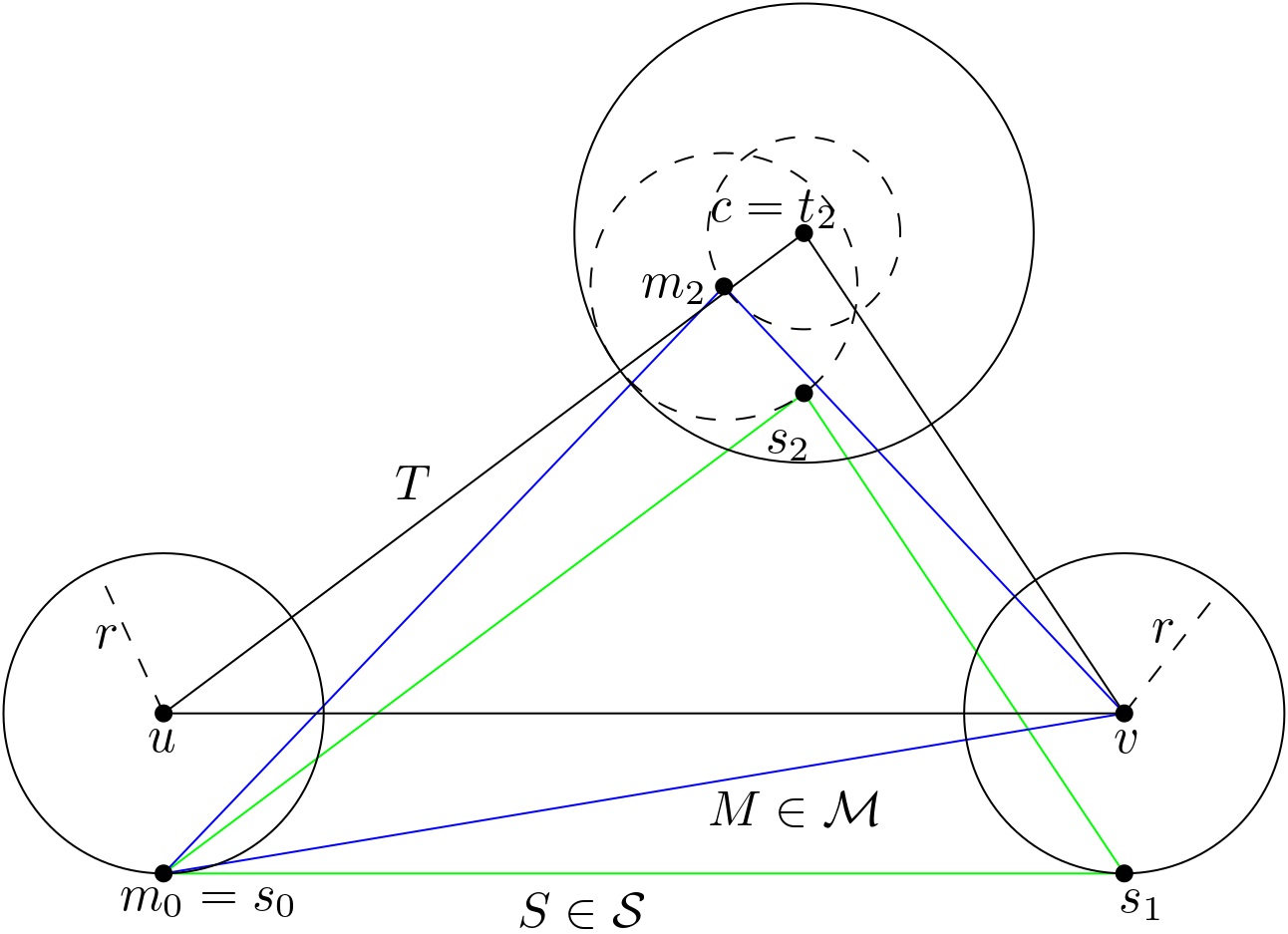}
      \caption{Replication Spanner}
      \label{fig:replication_spanner}
   \end{figure}

   Consider the Trivial Replication $T = R_{Triv}(P, u, v)$, and
   the replication machines $\mathcal{M} = R_{Mach}((p_1, p_0, p_2), v, u, r)$ and, for some 
   $M \in \mathcal{M}$, $\mathcal{S} = R_{Mach}(P, m_0, v, r)$ 
   (Figure \ref{fig:replication_spanner}).
   
   Observe that for any $S \in \mathcal{S}$, by the definition of Replication Spanner,
   $S \in R_{Span}(P, u, v, r)$.
   Observe that the center of the Replication Machine Circle of $\mathcal{S}$ is in 
   the Replication Machine Circle of $\mathcal{M}$.
   Therefore, $c$ is the center-of-centers of:
   \begin{align*}
      C(c, d(c, m_2) + d(m_2, s_2)) &= 
         C\left( c, r \frac{d(u, c)}{d(u, v)} + r\frac{d(v, c)}{d(u, v)} \right) \\
      &= C\left( c, r \frac{d(u, c) + d(v, c)}{d(u, v)} \right)
   \end{align*}
\end{appendixproof}

We call $C \left(c, r \frac{d(u, c) + d(v, c)}{d(u, v)} \right)$ 
the Replication Spanner Circle of $R_{Span}(P, u, v, r)$.

\section{Three-Robot Solution}
\label{sec:main_results}
In this section, we present the main contribution of this study: a solution for systems of three robots.
First, we show the optimal solution under rigid similarity, that is, we do not consider assignment (i.e. robot $i$ with initial position $r_i$ will assume the role of $p_i$ in the desired pattern).
Note that this is not necessarily the \textit{optimal} solution.
For systems of three robots, there are $3!$ possible assignments (permutations) of $P$ that could be optimal.
After presenting the solution for the trivial assignment, we demonstrate a simple method for choosing the correct assignment without testing all $3!=6$ possibilities.
Algorithm~\ref{alg:sol} produces a construction based entirely on geometric properties.
\begin{algorithm}
   \caption{Algorithm for robot $i$ in system with current positions $R$ to form pattern $P$}
   \label{alg:sol}
   \begin{algorithmic}[1]
      \Statex[0] \textit{// Let the perimeter of $P$ be $1$}
      \Statex[0] \textit{// indices are modulo 3}
      \State $t_i \gets $ point such that $\angle t_i r_{i+1} r_{i-1} = \angle p_i p_{i+1} p_{i-1}$ and $\angle r_{i+1} r_{i-1} t_i = \angle p_{i+1} p_{i-1} p_i$
      \State $r \gets d(r_i, t_i) ~ d(p_{i+1}, p_{i-1})$ \label{alg:sol:line:dist}
      \State $r_i $ moves $r$ toward $t_i$ \label{alg:sol:line:move}
   \end{algorithmic}
\end{algorithm}

\begin{lemmarep}
   \label{lemma:same_dist}
   For any system of robots with initial positions $R$ and any triangular pattern $P$,
   the distance $r$ computed in Algorithm~\ref{alg:sol} (Line~\ref{alg:sol:line:dist}) is the same for each robot.
\end{lemmarep}

Recall that for systems of three robots, all robots travel exactly the same distance.
We show in Lemma~\ref{lemma:same_dist} that Algorithm~\ref{alg:sol} satisfies this necessary condition.

\begin{appendixproof}
   For each robot $i$, note that $t_i$ is the Trivial Replication Point of $P$ on $r_{i+1}$ and $r_{i-1}$ (Figure~\ref{fig:construction_1}).
   Observe that $(r_0, r_1, t_0)$ is similar to $(t_2, r_1, r_2)$ by rotating around $r_1$ and uniformly scaling by some constant $k$, therefore
   $d(r_0, t_0) = k ~ d(r_2, t_2)$ and 
   $d(r_0, r_1) = k ~ d(r_1, t_2) \Rightarrow d(r_1, t_2) = \frac{d(r_0, r_1)}{k}$.
   Therefore, 
   \begin{align*}
      d(r_0, t_0) ~ d(p_1, p_2) &= d(r_0, t_0) ~ \frac{d(r_1, r_2)}{d(r_0, r_1) + d(r_1, r_2) + d(r_2, r_0)} \\
      &= \frac{k ~ d(r_2, t_2) ~ d(r_0, r_1)}{ k ~ (d(r_0, r_1) + d(r_1, r_2) + d(r_2, r_0))} \\
      &= d(r_2, t_2) ~ d(p_0, p_1) \\
   \end{align*}
   Using a similar argument, observe that 
   $d(r_0, t_0) ~ d(p_1, p_2) = d(r_1, t_1) ~ d(p_2, p_0)$.

   \begin{figure*}[ht!]
      \centering
      \begin{subfigure}[t]{0.49\textwidth}
         \centering
         \includegraphics[width=0.7\textwidth]{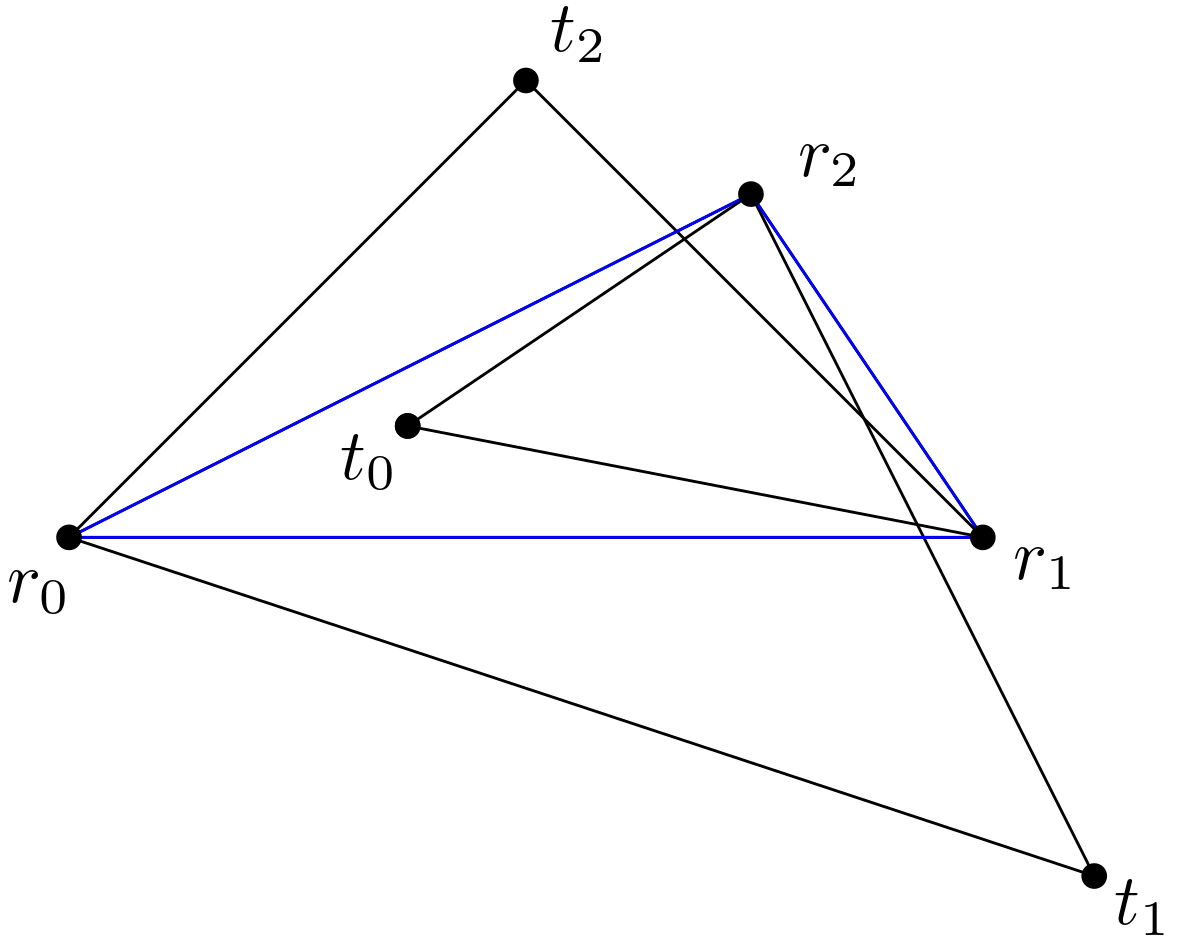}
         \caption{Algorithm~\ref{alg:sol} trivially replicates the desired pattern on each pair of points in the initial configuration of the system.}
         \label{fig:construction_1}
      \end{subfigure}
      ~ 
      \begin{subfigure}[t]{0.49\textwidth}
         \centering
         \includegraphics[width=\textwidth]{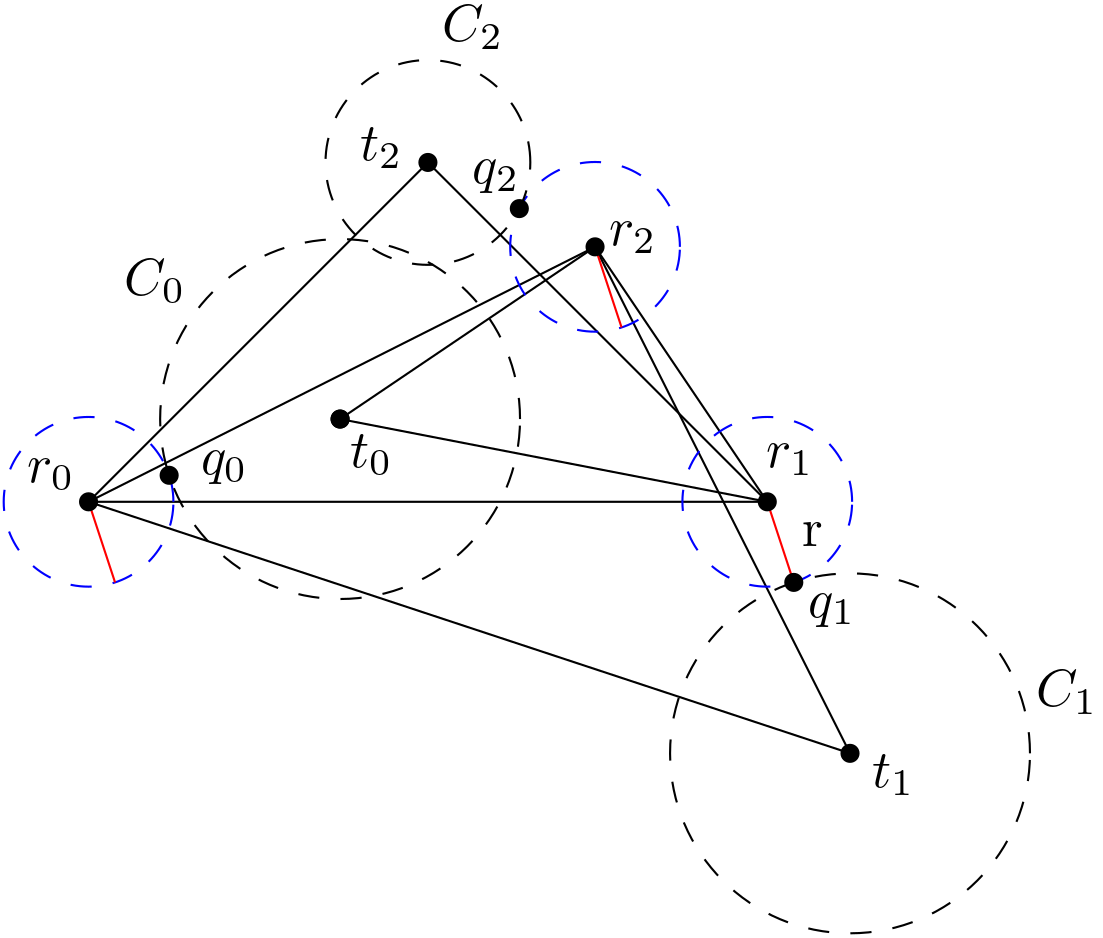}
         \caption{The Replication Machine Circles from the Replication Spanner of $P$ on each pair of points in the initial configuration of the system.}
         \label{fig:construction_2}
      \end{subfigure}
      \caption[Geometric Construction Replications]{}
      \label{fig:construction}
   \end{figure*}
\end{appendixproof}

\begin{theoremrep}
   For any system of robots with initial positions $R$ and triangular pattern $P$ with perimeter $1$, let $Q$ be the positions that robots move to after running Algorithm~\ref{alg:sol}. 
   Then $Q$ is a valid solution. In other words, $Q$ is similar to $P$.
\end{theoremrep}

\begin{appendixproof}
   Let $C_i$ be the Replication Spanner Circle of $P$ on
   $(C(r_{i+1}, r), C(r_{i-1}, r))$
   (Figure~\ref{fig:construction_2}).
   Observe that $C_i$ is centered at $t_i$ and has radius:
   \begin{align*}
      r^\prime &= r ~ \frac{d(t_i, r_{i+1}) + d(t_i, r_{i-1})}{d(r_{i+1}, r_{i-1})} \\
      &= d(r_i, t_i) ~ d(p_{i+1}, p_{i-1}) ~ \frac{d(t_i, r_{i+1}) + d(t_i, r_{i-1})}{d(r_{i+1}, r_{i-1})} \\
      &= d(r_i, t_i) \frac{d(r_{i+1}, r_{i-1})}{d(r_i, r_{i+1}) + d(r_{i+1}, r_{i-1}) + d(r_{i-1}, r_i)} \\
        &\phantom{=} ~ \frac{d(t_i, r_{i+1}) + d(t_i, r_{i-1})}{d(r_{i+1}, r_{i-1})} \\
      &= d(r_i, t_i) \frac{d(t_i, r_{i+1}) + d(t_i, r_{i-1})}{d(r_i, r_{i+1}) + d(r_{i+1}, r_{i-1}) + d(r_{i-1}, r_i)} \\
      &= d(r_i, t_i) ~ (d(p_i, p_{i+1}) + d(p_i, p_{i-1}))
   \end{align*}
   Then, notice that:
   \begin{align*}
      r + r^\prime &= d(r_i, t_i) ~ d(p_{i+1}, p_{i-1}) \\
        &\phantom{=} ~ + d(r_i, t_i) ~ (d(p_i, p_{i+1}) + d(p_i, p_{i-1})) \\
      &= d(r_i, t_i) ~ (d(p_{i+1}, p_{i-1}) + d(p_i, p_{i+1}) \\
        &\phantom{=} ~ + d(p_i, p_{i-1})) \\
      &= d(r_i, t_i)
   \end{align*}
   Therefore each robot moves to the single point of intersection of the Replication Machine Circle, and the circle $C(r_i, r)$ (Figure~\ref{fig:construction_2}).

   Without loss of generality, consider $q_2$.
   Observe that since $q_2 \in C_2$, there must exist a unique pair of points 
   $q^\prime_0 \in C(r_0, r)$ and 
   $q^\prime_1 \in C(r_1, r)$ such that 
   $Q^\prime = (q^\prime_0, q^\prime_1, q_2) \esim P$. 
   This fact follows from the definition of Replication Spanner (Section~\ref{sec:replication}).
   Now suppose, for sake of contradiction, that $q^\prime_0 \neq q_0$.
   This implies that $q^\prime_0 \notin C_0$ and thus
   $Q^\prime \notin R_{Span}((p_1, p_2, p_0), r_1, r_2, r)$
   (Figure~\ref{fig:construction_valid}).
   Thus, by the definition of Replication Spanner,
   $Q^\prime$ cannot be rigidly similar to $P$, a contradiction.
   By a similar argument, observe that $q_1 = q^\prime_1$.

   \begin{figure}[!ht]
      \centering 
      \includegraphics[width=\columnwidth]{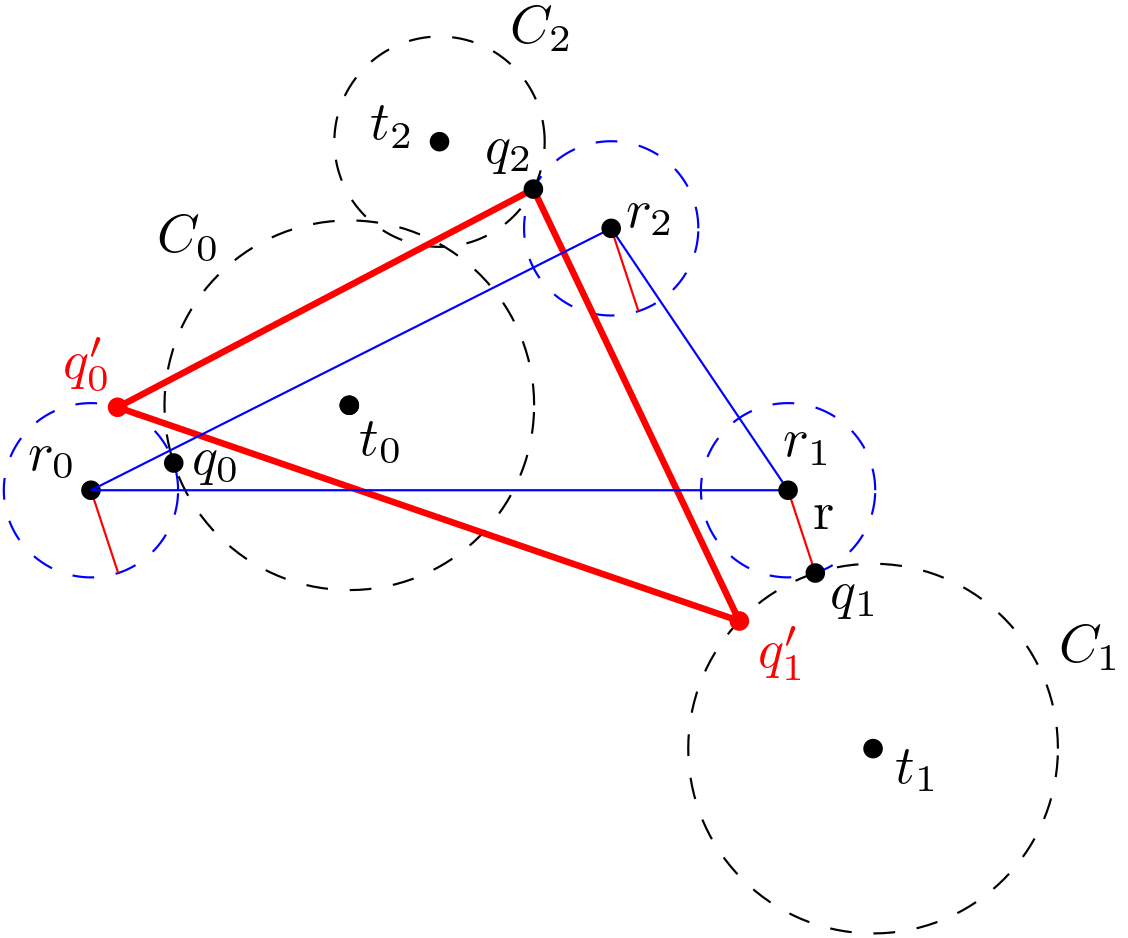}
      \caption[An invalid solution]{
         Since $q_2 \in C_2$, there must exist a pattern $Q^\prime \esim P$ such that $q^\prime_0 \in C(r_0, r)$ and $q^\prime_1 \in C(r_1, r)$. Observe this can only be true if $q^\prime_0 = q_0$ and $q^\prime_1 = q_1$.
      }
      \label{fig:construction_valid}
   \end{figure}
\end{appendixproof}

Algorithm~\ref{alg:sol} computes a valid solution such that all robots move the same distance.
These conditions are necessary for any optimal solution, although not sufficient.
We show in Theorem~\ref{theorem:geometric} that the solution Algorithm~\ref{alg:sol} produces is optimal.

\begin{theoremrep}
   \label{theorem:geometric}
   For any system of robots with initial positions $R$ and triangular pattern $P$, let $Q$ be the positions that each robot moves to after running Algorithm~\ref{alg:sol}, then $Q$ is an optimal formation under rigid similarity.
\end{theoremrep}

\begin{appendixproof}
   \begin{figure*}[ht!]
      \centering
      \begin{subfigure}[t]{0.49\textwidth}
         \centering
         \includegraphics[width=\textwidth]{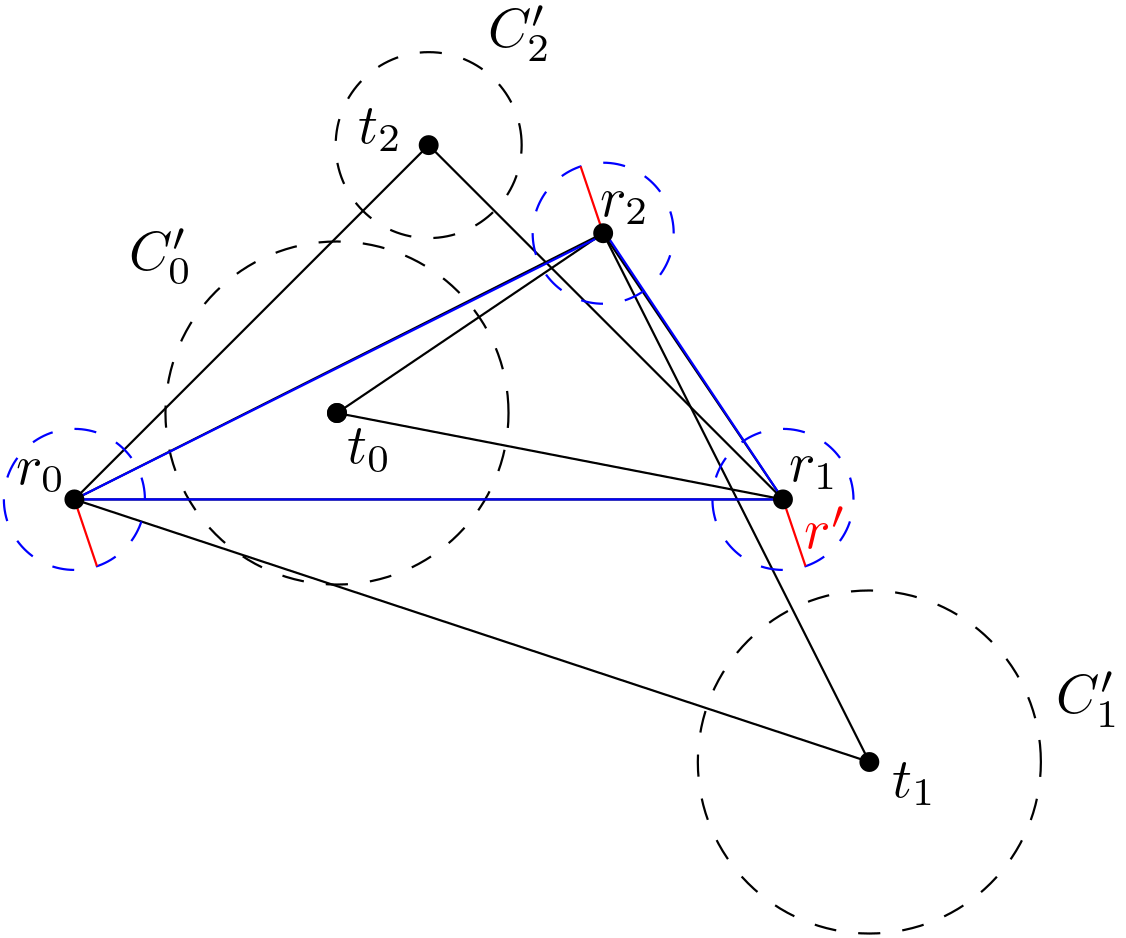}
         \caption{
            The distance $r^\prime$ results in an invalid pattern. 
            The robots cannot possibly form the desired pattern because, for each robot, all valid patterns that can be formed by the other two robots lie on a circle further than distance $r$ from the robot.
         }
         \label{fig:construction_invalid}
      \end{subfigure}
      ~ 
      \begin{subfigure}[t]{0.49\textwidth}
         \centering
         \includegraphics[width=\textwidth]{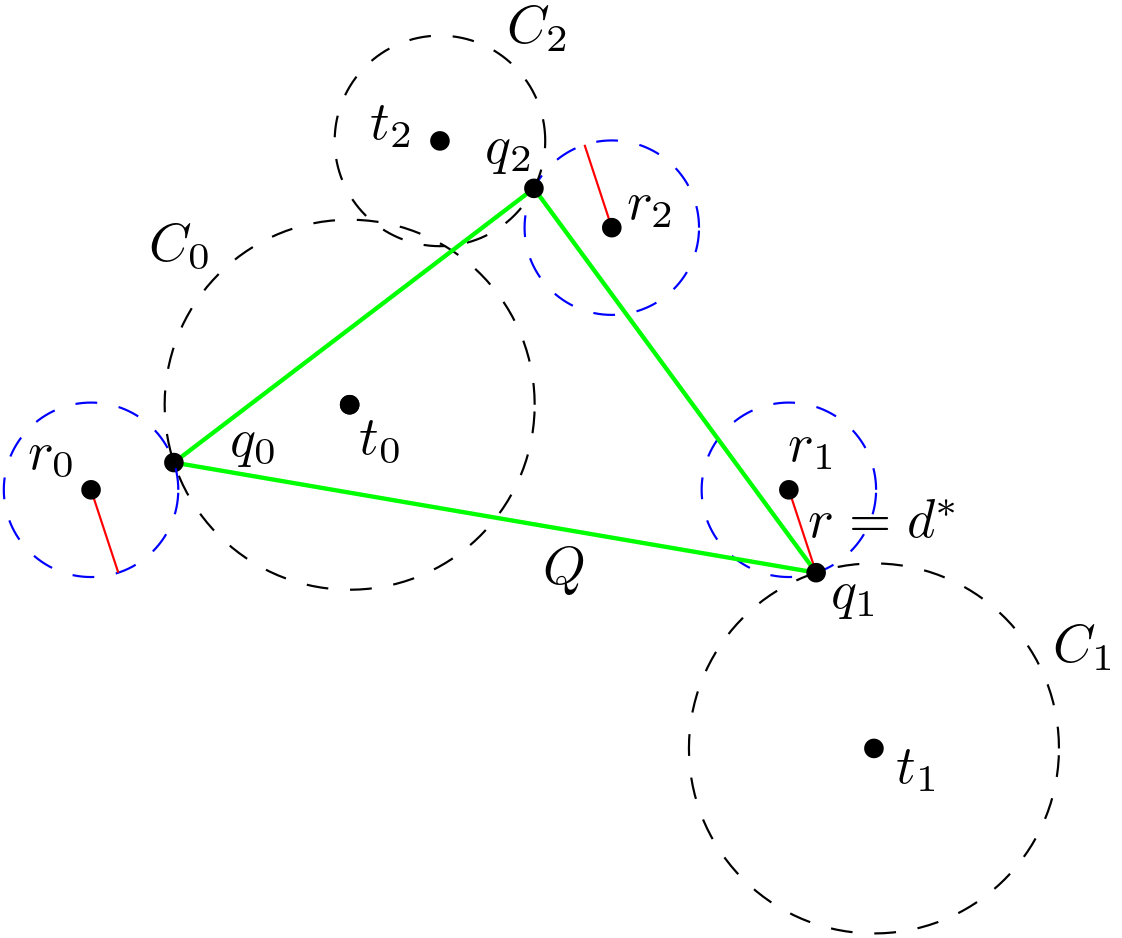}
         \caption{
            The distance $r$ is a valid value. 
            All robots are moving the same distance and, for each robot, the point of intersection lies on the appropriate Replication Spanner Circle.
         }
         \label{fig:construction_optimal}
      \end{subfigure}
      \caption[Geometric Construction - Final Steps]{}
      \label{fig:construction_radius}
   \end{figure*}

   Let $r^\prime$ be some radius arbitrarily smaller than $r$
   (Figure~\ref{fig:construction_invalid}).
   Without loss of generality, consider robot $0$.
   All rigidly similar patterns to $P$ that with $p_0 \in C(r_1, r^\prime)$
   lie outside of $C(r_1, r^\prime)$ and $C(r_2, r^\prime)$.
   This means that, in order to form a valid pattern, robots $2$ and $3$ would need to travel a distance greater than $r^\prime$.
\end{appendixproof}

\Subsection{Optimal Pattern Formation by Three Robots}
In order to prove Algorithm~\ref{alg:sol} is optimal, we assumed that robots move directly to their computed destinations.
In Section~\ref{sec:intro}, though, we discussed models where each robot executes \textit{look}, \textit{compute}, \textit{move} cycles.
In other words, we want to consider systems in which robots move a small distance $\epsilon$ toward their target, then re-compute the solution based on the new system state.
In this section, we show that our solution is valid for models with oblivious robots.

Consider a modification of Algorithm~\ref{alg:sol}, where instead of moving incrementally toward (rather than directly to) their destinations by replacing 
line~\ref{alg:sol:line:move} with:
$$
r_i \gets \text{ moves } min(r, \epsilon) \text{ toward } t_i
$$

\begin{theorem}
   \label{theorem:oblivious}
   Let $f_i(t)$ denote the position of robot $i$ at time $t$.
   For any $\epsilon > 0$, let $Q^t$ be the solution computed at time $t$.
   Then, $Q^t = Q^{t+1}$.
\end{theorem}

\begin{appendixproof}
   Recall that all robots move at the same speed.
   First, note that $d(f_i(t), q^{t+1}_i) \geq d(f_i(t), q^t_i)$, 
   or else $Q^t$ would not be optimal for time $t$. 
   Also, notice that $d(f_i(t+1), q^{t+1}_i) \leq d(f_i(t+1), q^t_i)$, or else
   $Q^t$ is a better solution than $Q^{t+1}$, which is a contradiction to the assumption that $Q^{t+1}$ is optimal.
   The only point where both of these conditions are satisfied is $q^{t+1} = q^t$ 
   (Figure \ref{fig:oblivious}).

   \begin{figure}[!ht]
      \centering
      \includegraphics[width=\columnwidth]{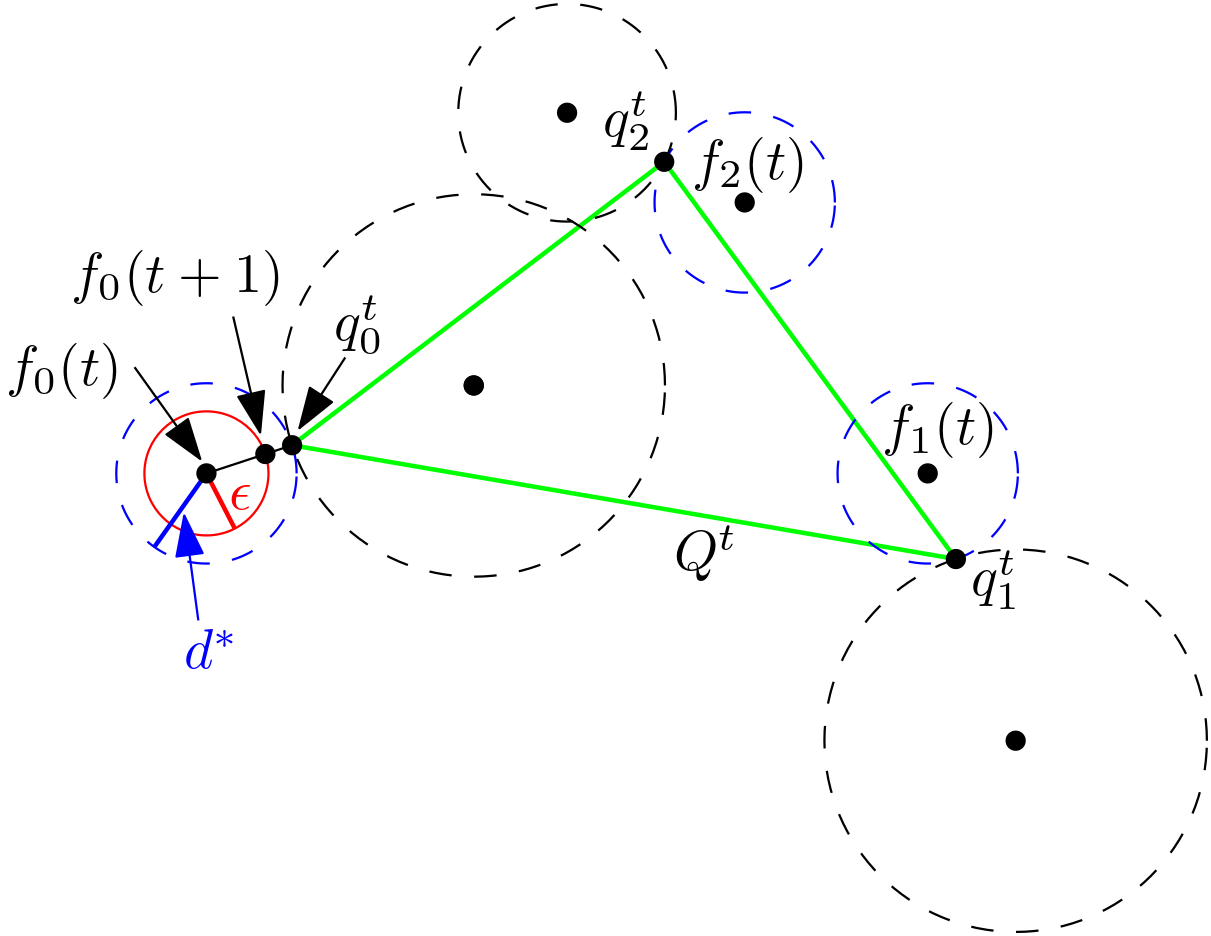}
      \caption[Robots can be oblivious]{
         At time $t+1$, $q^{t+1}_0$ must be on or outside $C(f_0(t), d^*)$ \textit{and} less than or equal to $d(f_i(t+1), q^t_0)$. The only place where both these conditions are met is $q^t_0$.
      }
      \label{fig:oblivious}
   \end{figure}
\end{appendixproof}

\Subsection{Assignment}
The geometric construction provides a solution under rigid similarity only and therefore does not consider different assignments (permutations) of the desired pattern.
In order to find the globally optimal solution, the geometric construction must be considered for all permutations of $P$.
In this section, we present a simple method for choosing the optimal assignment without testing all $3!=6$ possibilities.


\begin{theoremrep}
   \label{theorem:assignment}
   Consider a system of robots with initial positions 
   $R = (r_0, r_1, r_2)$, a pattern $P = (p_0, p_1, p_2)$,
   and $d(r_0, r_1) \leq d(r_1, r_2) \leq d(r_2, r_0)$.
   Then $P$ is an optimal assignment for $R$ if and only if
   $d(p_0, p_1) \leq d(p_1, p_2) \leq d(p_2, p_0)$.
\end{theoremrep}

\begin{appendixproof}
   Let $P^\prime = (p^\prime_0, p^\prime_2, p^\prime_1)$ 
   and suppose both $P$ and $P^\prime$ have perimeter $1$.
   First, suppose by contradiction that $P$ is optimal and, without loss of generality, that $d(p_1, p_2) > d(p_0, p_2)$.
   Consider the solution for this assignment $r = d(r_0, t) ~ d(p_1, p_2)$
   where $t$ is the Trivial Replication Point of $P$ on $(r_1, r_2)$.
   Now let $t^\prime$ be Trivial Replication Point of 
   $(p^\prime_0, p^\prime_2, p^\prime_1)$ on $(r_1, r_2)$
   with the solution 
   \begin{align*}
      r^\prime &= d(r_1, t^\prime_1) ~ d(p^\prime_0, p^\prime_2) \\
               &= d(r_1, t^\prime_1) ~ d(p_0, p_2)
   \end{align*}

   Since $d(p_1, p_2) > d(p_0, p_2)$ and $d(r_0, r_2) \leq d(r_0, r_1)$ 
   observe that $d(r_0, t^\prime) \leq d(r_0, t)$ (Figure~\ref{fig:assignment}).
   Therefore $r^\prime \leq r$ and $P^\prime$ is a better assignment.

   \begin{figure}[!h]
      \centering 
      \includegraphics[width=\columnwidth]{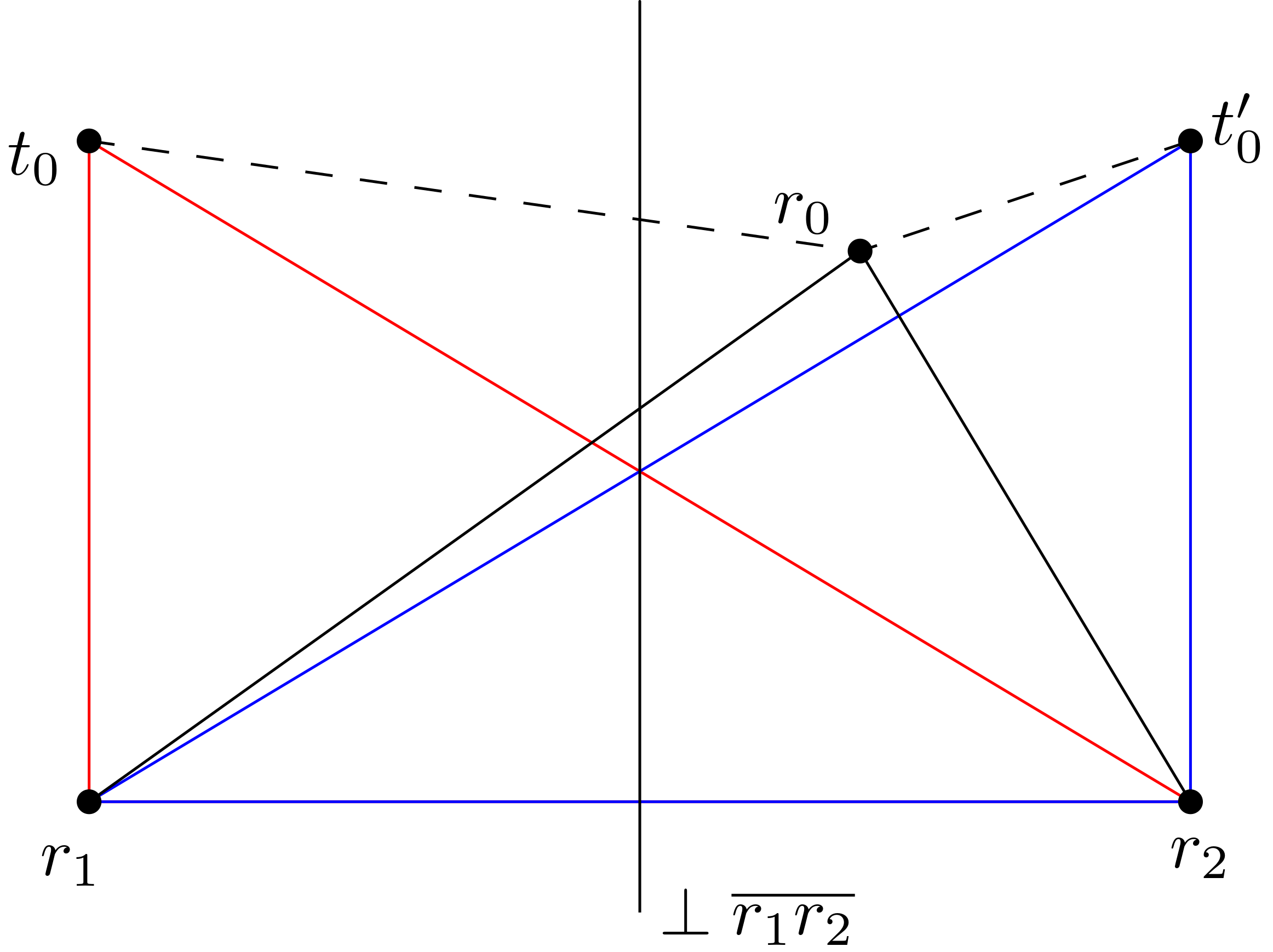}
      \caption[Assignment]{
         Notice that $t^\prime$ is closer to $r_0$ than $t$. 
         Since the solution for assignments $P$ and $P^\prime$ are proportional to $d(t, r_0)$ and $d(t^\prime, r_0)$ respectively, then $P^\prime$ is clearly a better assignment than $P$.
      }
      \label{fig:assignment}
   \end{figure}
\end{appendixproof}

Observe that, for any triangle $P$,
$d(p_0, p_1) \leq d(p_1, p_2) \leq d(p_2, p_0)$ if and only if 
$\angle p_{i-1} p_0 p_{i+1} \leq \angle p_0 p_{i+1} p_{i-1} \leq \angle p_{i+1} p_{i-1} p_0$.
Theorem~\ref{theorem:assignment} indicates that the optimal formation can be obtained by first sorting $R$ and $P$ by their angles (or side lengths), and then running Algorithm~\ref{alg:sol}.

\begin{correp}
   \label{cor:solution}
   For a system of robots with initial positions $R = (r_0, r_1, r_2)$, 
   such that $d(r_0, r_1) \leq d(r_1, r_2) \leq d(r_2, r_0)$
   and a pattern $P = (p_0, p_1, p_2)$ such that 
   $d(p_0, p_1) \leq d(p_1, p_2) \leq d(p_2, p_0)$
   let $Q$ be the positions that robots move to after running Algorithm~\ref{alg:sol}.
   Then $Q$ is an optimal formation.
\end{correp}

\begin{inlineproof}
   Follows from Theorems~\ref{theorem:geometric} and~\ref{theorem:assignment}.
\end{inlineproof}

\section{Triangle Metric}
\label{sec:metric}
In this section, we introduce a metric for comparing triangles inspired by the solution for systems of three robots presented in Section~\ref{sec:main_results}.
Let $d^*(A, B)$ be the optimal distance that robots with initial positions $A$ need to form $B$.
This distance can also be interpreted as a distance between the triangles $A$ and $B$.
$d^*$ is not a valid metric by itself, though.
In particular, since $d^*$ depends on the position and size of the first argument only, it is not symmetric, or 
$d^*(A, B) \neq d^*(B, A)$.
In order to enforce symmetry, our metric should be invariant to translation, rotation, uniform scaling, reflection, and permutation of both $A$ and $B$.

\begin{lemmarep}
   Let $\alpha$ and $\beta$ the ordered sequences of interior angles of two triangles.
   Then $\tau$ is a valid metric for comparing the triangles:
   \begin{align*}
      \tau^2(\alpha, \beta) &= \frac{sin^2(\alpha_1)}{sin^2(\alpha_2)} 
         + \frac{sin^2(\beta_1)}{sin^2(\beta_2)} \\
         &\phantom{=} ~ - 2\frac{sin(\alpha_1) ~ sin(\beta_1)}{sin(\alpha_2) ~ sin(\beta_2)}
         ~ cos(\alpha_0 - \beta_0)
   \end{align*}
\end{lemmarep}

\begin{appendixproof}
   Let $B$ be the optimal assignment for a triangle $A$
   (Section~\ref{sec:main_results}).
   Now consider the Trivial Replications $T$ and $T^\prime$ of $A$ and $B$, respectively, on $((0, 0), (1, 0))$ (Figure~\ref{fig:metric}).
   
   \begin{figure}[!ht]
      \centering 
      \includegraphics[width=\columnwidth]{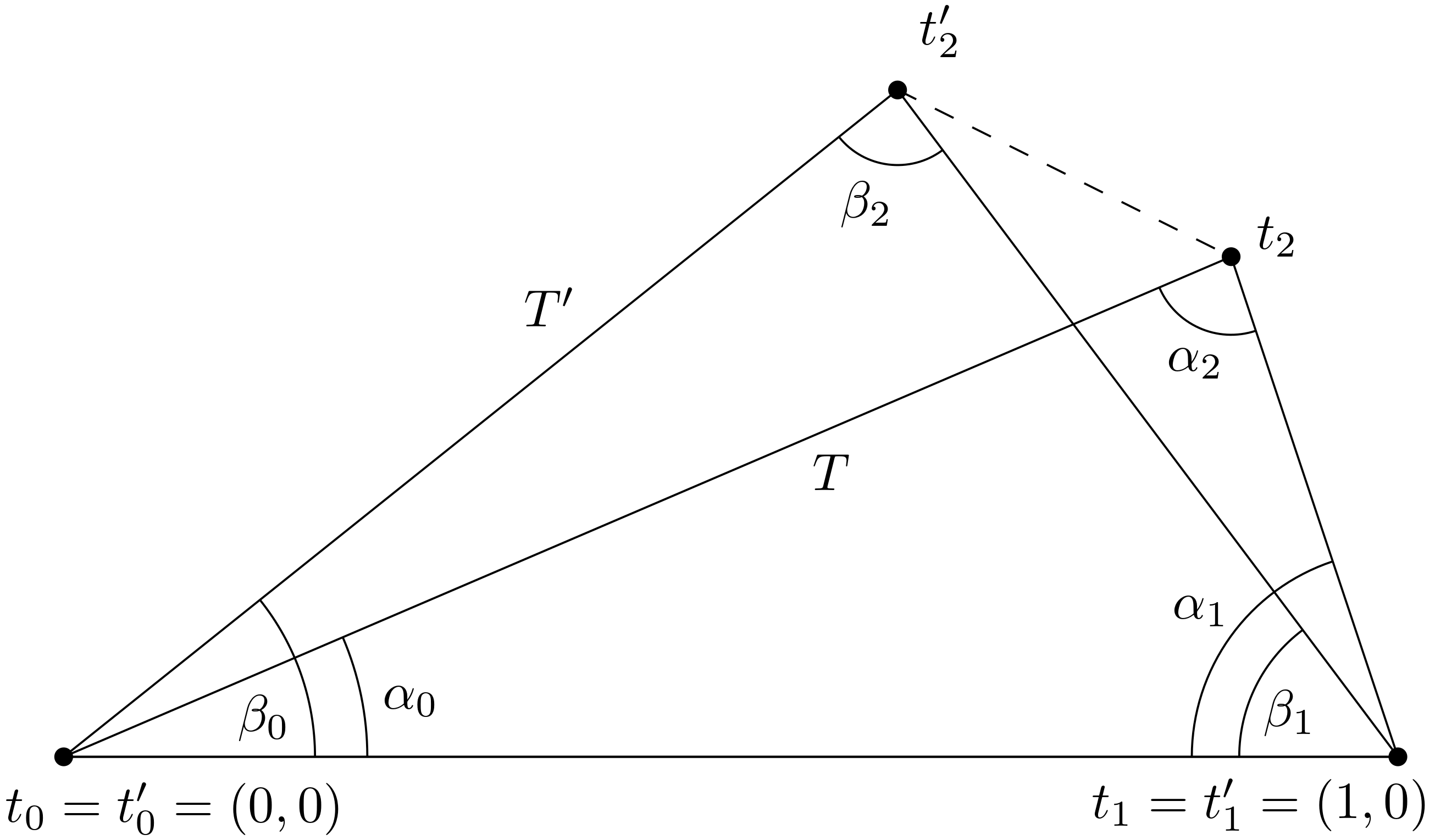}
      \caption[Triangle Metric]{
         The metric $\tau$ is defined as the distance between the Trivial Replication Points of the two triangles on $((0, 0), (1, 0))$.
      }
      \label{fig:metric}
   \end{figure}

   Observe the resemblance between $d^*(T, B)$ and $d^*(T^\prime, A)$:
   \begin{align*}
      d^*(T, B) &= d(t_2, t^\prime_2) ~ d(b_0, b_1) \\
         &= d(t_2, t^\prime_2) ~ \frac{d(t_0, t_1)}{d(t_0,t_1) + d(t_1,t_2) + d(t_2,t_0)} \\
         &= \frac{d(t_2, t^\prime_2)}{d(t_0,t_1) + d(t_1,t_2) + d(t_2,t_0)} \\
      \: & \\
      d^*(T^\prime, A) &= d(t_2, t^\prime_2) ~ 
         \frac{d(t^\prime_0, t^\prime_1)}{d(t^\prime_0,t^\prime_1) + d(t^\prime_1,t^\prime_2) + d(t^\prime_2,t^\prime_0)} \\
         &= \frac{d(t_2, t^\prime_2)}{d(t^\prime_0,t^\prime_1) + d(t^\prime_1,t^\prime_2) + d(t^\prime_2,t^\prime_0)}
   \end{align*}
   Note that both $d^*(T, B)$ and $d^*(T^\prime, A)$ 
   are proportional to the distance $d(t_2, t^\prime_2)$ between their Trivial Replication Points.
   Our metric, $\tau$ is this distance.
   Given two triangles, we can compute the $\tau$-distance between them 
   (Figure~\ref{fig:metric}).
   First, observe that by the law of sines, 
   $\frac{d(t_0, t_2)}{sin(\alpha_1)} = \frac{1}{sin(\alpha_2)}$, so
   $d(t_0, t_2) = \frac{sin(\alpha_1)}{sin(\alpha_2)}$.
   Similarly, $d(t^\prime_0, t^\prime_2) = \frac{sin(\beta_1)}{sin(\beta_2)}$
   Then, by the law of cosines:
   \begin{align*}
      &\tau(A, B) = d(t_2, t^\prime_2) \\
         &= \sqrt{ 
            d^2(t_0, t_2) + d^2(t^\prime_0, t^\prime_2) 
            - 2 d(t_0, t_2) ~ d(t^\prime_0, t^\prime_2) ~ cos(\alpha_0 - \beta_0) 
         } \\
         &= \sqrt{
            \frac{sin^2(\alpha_1)}{sin^2(\alpha_2)} 
            + \frac{sin^2(\beta_1)}{sin^2(\beta_2)}
            - 2\frac{sin(\alpha_1) ~ sin(\beta_1)}{sin(\alpha_2) ~ sin(\beta_2)}
            ~ cos(\alpha_0 - \beta_0)  
         }
   \end{align*}
   where $\alpha$ and $\beta$ are the ordered interior angles of $A$ and $B$, 
   respectively (i.e. $\alpha_0 \leq \alpha_1 \leq \alpha_2$
   and $\beta_0 \leq \beta_1 \leq \beta_2$).
   
   Now consider three triangles $A$, $B$, and $C$.
   By definition, $\tau$ is a metric if and only if the following four properties are satisfied:
   \begin{enumerate}[topsep=0pt,itemsep=-1ex,partopsep=1ex,parsep=1ex]
      \item $\tau(A, B) \geq 0$      
            \hspace*{\fill} non-negativity
      \item $\tau(A, B) = 0 \Leftrightarrow A \sim B$ 
            \hspace*{\fill} identity of indiscernibles
      \item $\tau(A, B) = \tau(B, A)$ 
            \hspace*{\fill} symmetry
      \item $\tau(A, C) \leq \tau(A, B) + \tau(B, C)$
            \hspace*{\fill} triangle inequality
   \end{enumerate}

   Note that we define two patterns to be equal under our metric if they are similar to each other, or $A \sim B$.
   Clearly, $\tau$ satisfies non-negativity, since it is a distance between two points on the plane.
   It is also clear to see algebraically that the symmetry and identity of indiscernibles properties are also satisfied.
   Finally, $\tau$ satisfies the triangle inequality by the definition of Euclidian Distance.
   Observe in Figure~\ref{fig:triangle_neq} that 
   $\tau(A, C) \leq \tau(A, B) + \tau(B, C)$.
   
   \begin{figure}[!ht]
      \centering 
      \includegraphics[width=\columnwidth]{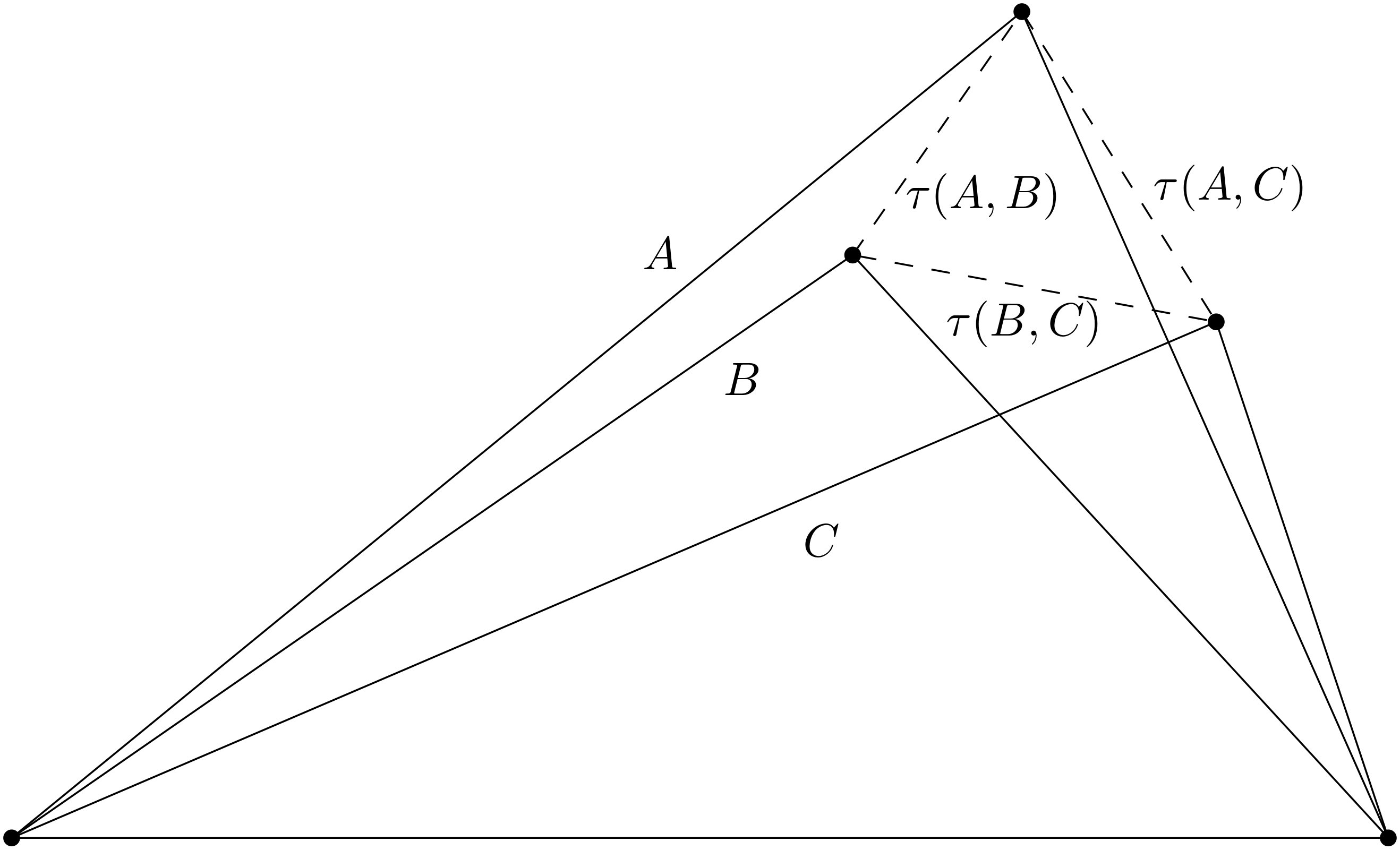}
      \caption[Triangle Metric: Triangle Inequality]{
         The triangle inequality holds since $\tau$ is a Euclidian distance, by definition. Observe that $\tau(A, C) \leq \tau(A, B) + \tau(B, C)$.
      }
      \label{fig:triangle_neq}
   \end{figure}
\end{appendixproof}

The $\tau$-distance between two triangles, defined by their angles, is a measure of similarity between them.
Two triangles, $A$ and $B$ are similar when $\tau(A, B) = 0$.
If $\tau(A, B) < \tau(A, C)$ this indicates that $B$ is \textit{more similar} to $A$ than $C$ is.
In other words, a system of robots with an initial formation of $A$ would need to travel further to form $C$ than it would move in order to form $B$.

\section{Arising Geometric Properties}
\label{sec:properties}

In this section, we present some interesting properties of systems of three robots and the patterns they can form.

\Subsection{Focal Point}
One interesting property that emerges for every system of three robots forming any arbitrary pattern is that all three of their paths can be characterized by a single point on the plane.

\begin{theoremrep}
   For systems of three robots and any optimal formation, there exists a point that all robots move either directly toward or directly away from.
\end{theoremrep} 

\begin{appendixproof}
   Consider a triangle $(a, b, c)$, and a pattern $P$. 
   Then, consider the three triangles rigidly similar to $P$ that share two vertices with $(a, b, c)$, namely $(a^\prime, b, c)$, $(a, b^\prime, c)$, and 
   $(a, b, c^\prime)$ (Figure~\ref{fig:focal_point}).
   Observe that this is equivalent to the solution for a system of robots with initial positions
   $R = (a, b, c)$ to form the pattern $P \sim (a, b, c^\prime)$.
   To show that every robot moves toward a single point, we must show that there exists some point $f$ which lies on $\overleftrightarrow{aa^\prime}$, $\overleftrightarrow{bb^\prime}$, and $\overleftrightarrow{cc^\prime}$.

   Let $\alpha = \angle bac^\prime$, 
   $\beta = \angle abc^\prime$,
   $\gamma = \angle ac^\prime c$, and 
   $\gamma^\prime = \angle bb^\prime c$.
   Note that $\gamma + \gamma^\prime = \angle ac^\prime c$.

   Observe that, by the definition of rotation and similar triangles,
   $\angle abb^\prime = \gamma$
   and $\angle baa^\prime = \gamma^\prime$.

   Also observe that
   \begin{align*}
   \angle b f c &= \pi - (\beta + \gamma + \gamma^\prime) \\
   &= \pi - (\beta + \angle ac^\prime b) \\
   &= \alpha
   \end{align*}

   Similarly, observe that $\angle afc = \beta$.

   Finally, $\gamma + \gamma^\prime + \alpha + \beta = \pi$.
   Therefore, $afb$ is a triangle and $f$ must be a single point.

   \begin{figure}[!ht]
      \centering
      \includegraphics[width=\columnwidth]{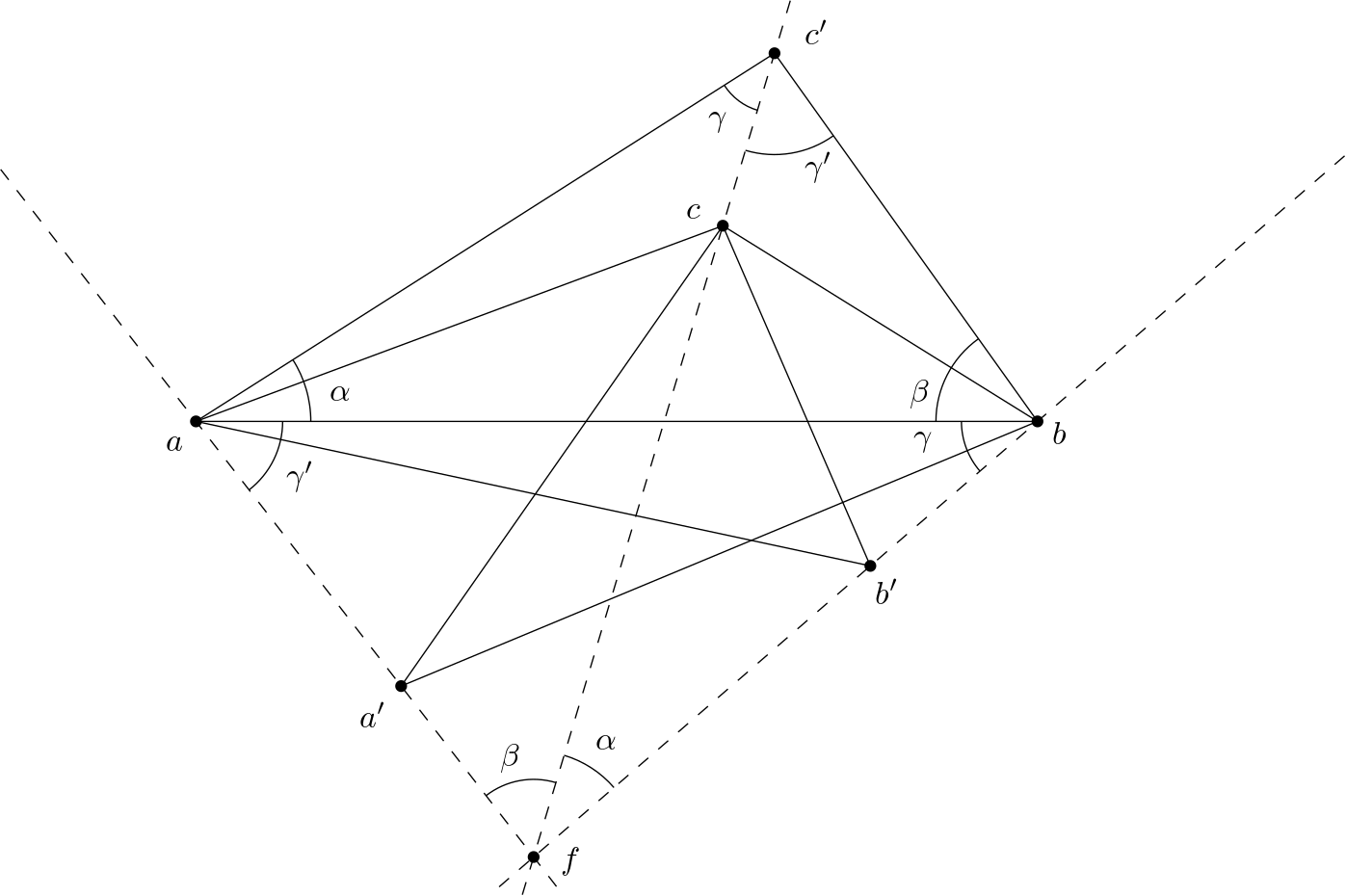}
      \caption[Focal Point]{All three robots move toward a single point.}
      \label{fig:focal_point}
   \end{figure}
\end{appendixproof}

For any optimal pattern, we call this single point that robots move either directly to or from, the \textit{focal point}.

\Subsection{Constant Center-of-Mass}
For systems forming equilateral triangles, an even stronger property emerges.

\begin{lemmarep}
   Suppose $Q$ is an optimal formation for a system of three robots with initial positions $R$ to form an equilateral triangle.
   Then, the center of mass of $Q$ is equivalent to that of $R$.
   Furthermore, since robots move at the same speed, the system's center of mass is invariant with respect to time.
\end{lemmarep}

\begin{appendixproof}
   Given a system of three robots with initial positions $R$, let $P$ be the equilateral triangle pattern it must form.
   First observe that the center of mass of the system at $t=0$ is: 
   \begin{align*}
      c = \left( 
         \frac{x(r_0) + x(r_1) + x(r_2)}{3}, 
         \frac{y(r_0) + y(r_1) + y(r_2)}{3}
      \right)
   \end{align*}
   Let the focal point of $Q$ be the origin and suppose, without loss of generality, that 
   $x(r_0) = 0$ (i.e. robot $0$ is on the $y$-axis; Figure~\ref{fig:center_of_mass}). 
   Without loss of generality, assume robots $0$ and $2$ move together (either both toward or both away from the focal point). 
   Observe then, that robot $1$ moves in the opposite direction relative to the focal point.
   We can then compute the center of mass of the system after a single time-step 
   (each robot moves one unit of distance towards its destination):
   \begin{align*}
         c^\prime &= \left( 
            \frac{x(r_0) + (x(r_1) + cos \frac{\pi}{6}) + (x(r_2) - cos \frac{\pi}{6})}{3},\right. \\
         &\quad \quad \left. 
            \frac{(y(r_0) + 1) + (y(r_1) - sin \frac{\pi}{6}) + (y(r_2) - sin \frac{\pi}{6})}{3}
         \right) \\
         &= \left( 
            \frac{x(r_0) + x(r_1) + x(r_2)}{3}, \frac{y(r_0) + y(r_1) + y(r_2)}{3}
         \right) = c
   \end{align*}
   
   \begin{figure}[!ht]
      \centering
      \includegraphics[width=\columnwidth]{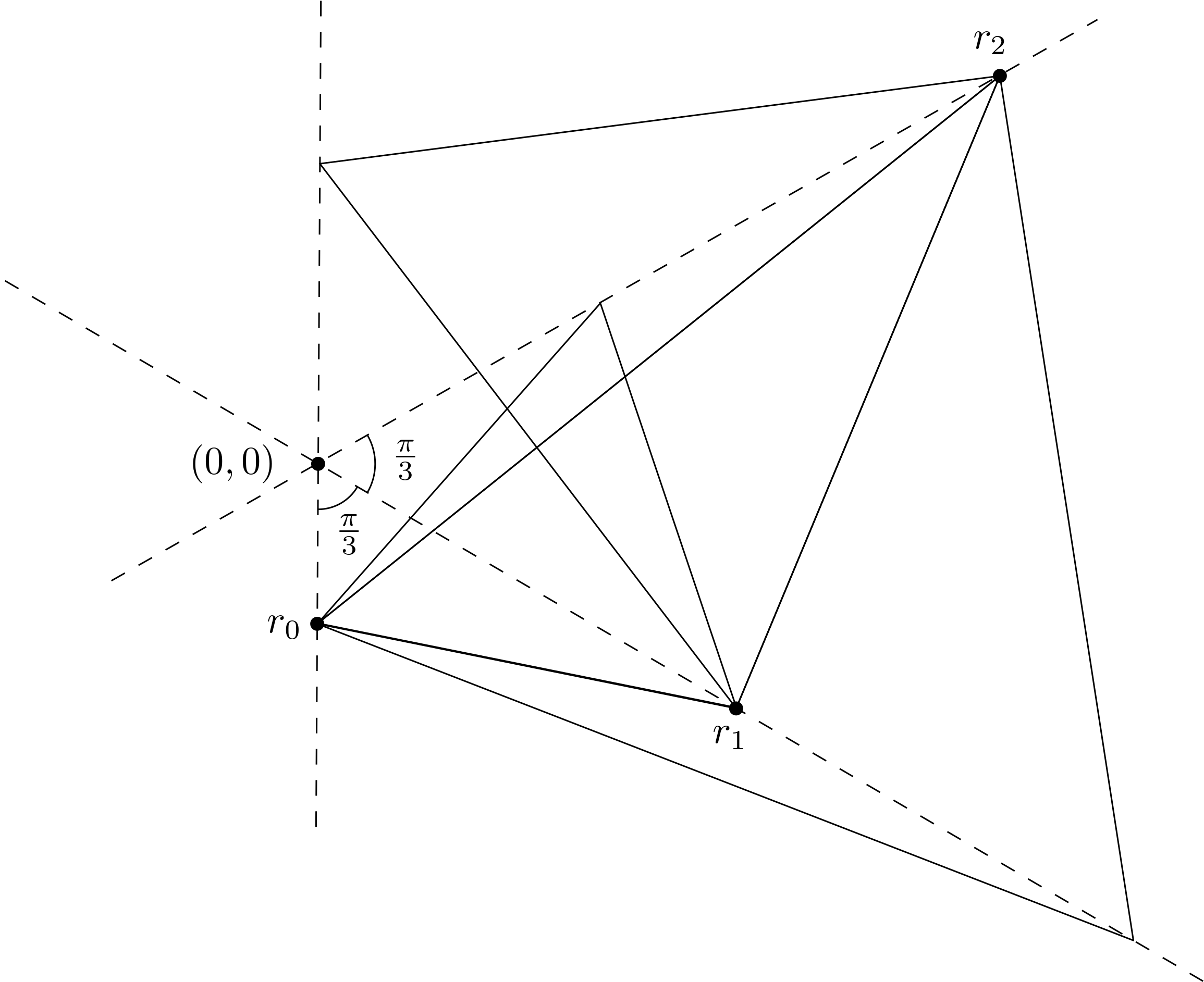}
      \caption[Constant center of mass for equilateral triangle formation]{
         A system of three robots move towards/away from the origin to form an equilateral triangle.
      }
      \label{fig:center_of_mass}
   \end{figure}

\end{appendixproof}

\section{Conclusion}
\label{sec:conclusion}
The main contribution of this study is an optimal solution for systems of three robots.
Systems of three robots are interesting because they have clear applications to systems of many robots.
Recall that, even in the general case, at least three robots must traverse the maximum distance, therefore it is a lower bound for the general case, that is, $d$ is the minimum optimal solution for all combinations of three robots and triangular sub patterns of $P$ with a prescribed assignment, or: 
$$
   d = \underset{p_i, p_j, p_k \in P}{\min}( 
      \underset{r_i, r_j, r_k \in R}{\max}(
         d^*((r_i, r_j, r_k), (p_i, p_j, p_k))
      )
   )
$$
Finding an upper bound on the solution is an area for future work.
A generalized Replication Machine tool might prove useful in finding the solution for systems of $n$ robots.

We are also interested in finding an algorithm for determining the optimal assignment in the general case.
It is clear that some assignments are infeasible.
For example, it makes intuitive sense that a robot's set of nearest neighbors in the initial configuration of the system should be close to that of final configuration.

Further work is also needed to understand under which models (see Section~\ref{sec:intro}) our solution (or some variant of it) is valid for.
For example, Algorithm~\ref{alg:sol} is only valid for synchronous models, where each robot starts its cycle at the same time (according to a global clock).
If the robots were asynchronous, they would compute optimal solutions for different initial configurations, since they would observe the current positions of robots at different times.

Finally, we plan to explore applications for the triangle metric introduced in Section~\ref{sec:metric}.
The metric provides a nice way to score, classify, or sort triangles based on their similarity to each other.
This has potential applications in computer vision, computational geometry, and of course, mobile robotics.

\paragraph{Acknowledgements.} This work was initiated at the 18th Routing Workshop which took place in Merida, Mexico from July 29 to August 02, 2019.
Research supported by PAPIIT grant IN 102117 from Universidad Nacional Aut\'onoma de M\'exico.
B.V.\, was partially supported by the European Union's Horizon 2020 research and innovation programme under the Marie Sk\l{}odowska-Curie grant agreement No 734922
and by the Austrian Science Fund within the collaborative DACH project \emph{Arrangements and Drawings} as FWF project \mbox{I 3340-N35}.

\pagebreak

\small
\bibliographystyle{abbrv}
\bibliography{main}

\end{document}